\documentclass[12pt,a4paper]{report}

\pdfoutput=1

\usepackage{amsthm,psfrag,graphicx}
\usepackage{subfigure}
\usepackage{amsmath,amssymb,amsfonts}
\usepackage{color}
\definecolor{darkblue}{rgb}{0.1,0.1,.7}
\usepackage[colorlinks, linkcolor=darkblue, citecolor=darkblue, urlcolor=darkblue, linktocpage]{hyperref} 
\usepackage[square, comma, compress,numbers]{natbib}
\usepackage[]{latexsym}
\usepackage{geometry}
\usepackage{amscd}
\usepackage[all,cmtip]{xy}
\usepackage{mathrsfs}
\usepackage[margin=10pt,font=small,labelfont=bf]{caption}
\numberwithin{equation}{chapter}

\newcommand{\reef}[1]{(\ref{#1})}

\newcommand{\bea}{\begin{eqnarray}}
\newcommand{\eea}{\end{eqnarray}}

\newcommand{\eps}{\epsilon}
\def\beq{\begin{equation}} 
\def\eeq{\end{equation}} 
\def\del {\partial} 
\def\nn{\nonumber} 
\def\bZ {\mathbb{Z}} 
\def\bR {\mathbb{R}} 
\def\calO {{\cal O}}

\def\bZ {\mathbb{Z}} 

\def\ge{\geqslant}
\def\le{\leqslant}

\newcommand{\be}[1]{\begin{equation} \centering \label{#1}}
\newcommand{\ee}{\end{equation}}

\begin{document}
\vspace*{-.6in} \thispagestyle{empty}
\begin{flushright}
CERN TH/2016-012
\end{flushright}
\vspace{1cm} 
{\Large
\begin{center}
{\bf EPFL Lectures on Conformal Field Theory\\ 
in $D\ge 3$ Dimensions}\\
\end{center}}
\vspace{1cm}
\begin{center}
{\bf Slava Rychkov$^{a,b,c}$}\\[2cm] 
{
$^{a}$ CERN, Theoretical Physics Department, Geneva, Switzerland\\[5pt]
$^{b}$ Laboratoire de Physique Th\'{e}orique (LPTENS),\\ Ecole Normale Sup\'{e}rieure, PSL Research University, Paris, France\\[5pt]
$^{c}$ Sorbonne Universit\'es, UPMC Univ Paris 06, Facult\'e de Physique, Paris, France
}
\\[1cm]
\vspace{1cm}
{\bf Abstract}
\end{center}

This is a writeup of lectures given at the EPFL Lausanne in the fall of 2012. The topics covered: physical foundations of conformal symmetry, conformal kinematics, radial quantization and the OPE, and a very basic introduction to conformal bootstrap.

\tableofcontents

\chapter*{Introduction}
\addcontentsline{toc}{chapter}{Introduction}

In the fall of 2012, I gave four lectures (3.5 hours each) on CFT in $D\ge 3$ to Master and PhD students at the EPFL Lausanne. A preliminary writeup was long available on my website. This version contains essentially the same material up to misprint and language corrections and a few additional comments. I also updated the references and added some historical remarks.

These lectures develop CFT from scratch. Lecture \ref{lecture1} deals with the physical foundations of conformal invariance. Lecture \ref{lecture2} considers constraints imposed by conformal symmetry on the correlation functions of local operators, presented using the ``projective null cone" (also known as the ``embedding") formalism. Lecture \ref{lecture3} is devoted to the radial quantization and the OPE. Lectures 1-3 are rather in-depth. Lecture \ref{lecture4} attempts an introduction to the conformal bootstrap, and is much more sketchy.  

For more details about the bootstrap, the reader may turn to the lectures by Sheer El-Showk \cite{Busan}
and David Simmons-Duffin \cite{TASI2015}. See also my recent lectures at DESY, Okinawa and Florence \cite{DESY,Okinawa,Florence}. Another useful resource are the lectures by Joshua Qualls \cite{Qualls:2015qjb}, which cover both $D=2$ and $D\ge 3$. Finally, I give some points of entry to the original literature after Lecture \ref{lecture4}.

\section*{Acknowledgement}

Conversations with Sergio Ferrara, Gerhard Mack, and Alexander Polyakov about the early CFT history are gratefully acknowledged; the possible fault of misrepresentation is fully on me. Many thanks to Georgios Karananas who took notes and TeX'ed them back in 2012, and to Pierre Yvernay and Francesco Riva for reporting the misprints. 
Please email me (\href{mailto:slava.rychkov@gmail.com}{slava.rychkov@gmail.com}) if you find more.


\chapter{Physical Foundations of Conformal Symmetry}
\label{lecture1}
\section{Fixed points}
Quantum Field Theory (QFT) is, in most general terms, the study of Renormalization Group (RG) flows, i.e. how the theory evolves from the Ultraviolet (UV) to the Infrared (IR) regimes: 
\begin{center}
\includegraphics[scale=0.7]{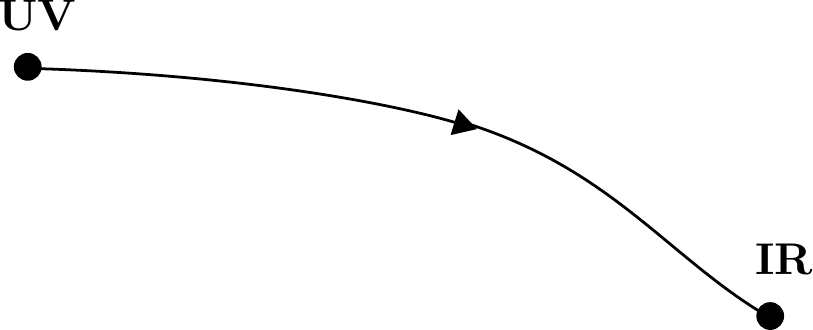}
\end{center}
One can ask which IR phases are possible. A priori, there are three possibilities:
\begin{itemize}
\item[A.]{a theory with a mass gap,}
\item[B.]{a theory with massless particles in the IR,}
\item[C.]{a Scale Invariant (SI) theory with a continuous spectrum.}
\end{itemize}
It is the last class that we will call CFT and will mostly study in these lectures.
But first let's look at some examples corresponding to these phases. 

Nonabelian Yang-Mills (YM) theory in $D=4$ dimensions belongs to type A:
\be{lagr1}
\mathscr L =-\frac{1}{4g^2}F^a_{\mu\nu}F^{a\mu\nu}\ .
\ee
The beta function is negative
\be{beta1}
\beta(g)\sim - c g^3+\mathcal{O}(g^5)\,, \quad c>0\, ,
\ee
so the theory becomes free at asymptotically high energies (the coupling goes to zero), while at low energies the coupling grows and the theory becomes nonperturbative at the scale
\beq
\Lambda_{IR}=\Lambda_{UV} \exp(-const./g_{UV}^2).
\eeq
The low energy spectrum will include the lightest ``glueball'', a scalar particle of mass $m\sim\Lambda_{IR}>0$, and some heavier stuff.
 
QED gives a trivial example of a theory with massless particles in the IR (i.e.~type B). For energies $E\ll m_e$, the mass of the electron, we are left with free photons. If the electron were massless, the IR behavior would still be Type B, but it would be reached very slowly (logarithmically)

A less trivial type B example is massless QCD with a small number $N_f$ of fermionic flavors. The UV theory exhibits invariance under $G=SU(N_f)_L\times SU(N_f)_R\times U(1)_B$. In the IR, the chiral symmetry breaks spontaneously $SU(N_f)_L\times SU(N_f)_R\rightarrow SU(N_f)_{diag}$, which results in a theory with massless Goldstone bosons. 

Examples of type C behavior exist among ``fixed point" theories for which the beta function vanishes at some point $g=g_*$, while the theory is still weakly coupled (i.e. we are still inside the perturbative regime):
\be{banks-zaks}
\beta(g_*)=0 \ , \ g_*\ll 1 \ .  
\ee
Consider again the YM theory with $N_c$  colors and $N_f$ fermions. The beta function at two loops is given by
\be{beta2}
\beta(g)=-\beta_0\frac{g^3}{16\pi^2}+\beta_1\frac{g^5}{(16\pi^2)^2}+ O (g^7) \ ,
\ee
with 
\be{def1}
\beta_0=\frac{11}{3}N_c-\frac 23 N_f \ \ \  \text{and} \ \ \  \beta_1\sim  O(N_c^2,N_c N_f) \ .
\ee
 Let us choose $N_c,N_f\gg 1$, in such a way that $\beta_0\sim O(1)>0$ (near cancelation). With this choice one can check that $\beta_1\sim O(N_c^2)>0$ (no cancellation). Writing
\be{beta3}
\beta(g)=-\frac{g^3}{16\pi^2}\left(1-\frac{\beta_1}{\beta_0}\frac{g^2}{16\pi^2}+\ldots \right) \ ,
\ee
we see that there is a zero at $g=g_*$ with $g^2_*/(16\pi^2)=O(1/N^2_c)$. Define $\lambda_*$ by
\be{lambda}
\lambda_*=\frac{N_cg^2_*}{16\pi^2}\sim O(1/N_c) \ ,
\ee 
The effects of the corrections from higher order terms are suppressed by powers of the coupling $\lambda_*$, so the perturbative expansion is trustworthy. This IR fixed point is known as the ``Banks-Zaks (BZ) fixed point".

For a finite $N_c$, depending on $N_f$, there will be the following cases:
\begin{itemize}
\item[1.]{For $N_f$ small described above, we will have chiral symmetry breaking (Type B)}
\item[2.]{In a range
$
N_{f,critical}<N_f<\frac{11}{2}N_c \ 
$
(the so called ``conformal window"), the theory possesses a Banks-Zaks fixed point (Type C)}
\item[3.]{For $N_f>11/2 N_c$, the theory is not asymptotically free, like massless QED. So IR behavior is asymptotically Type B.}
\end{itemize}

An IR behavior of a generic fixed point theory will be very much unlike standard QFT: the IR spectrum will be continuous and there will be no well-defined particles. When the theory flows, any gauge-invariant operator acquires an anomalous dimension, and when the fixed point is reached, this anomalous dimension ``freezes".  E.g. the anomalous dimension of $\overline\psi\psi$ at the BZ fixed point freezes at 
\be{anom1}
\gamma(g_*)=-\frac{g_*^2}{2\pi^2}\neq0 \ .
\ee
This means that this operator, and generically any operator, will have non-integer dimensions at the IR fixed point:
\be{corr1}
\Delta=\Delta_{free}+\gamma(g_*)  \ . 
\ee
This in turn implies that the spectrum of this theory will be continuous. 

To see why this is so, consider the two-point (2pt) function of an operator $\phi$. Inserting the complete basis of states, we can write:
\be{corr2}
\langle0\vert\phi(x)\phi(0)\vert0\rangle=\int_{F.C.}\frac{d^4p}{(2\pi)^4}e^{-ipx}\vert\langle0\vert\phi\vert p\rangle\vert^2\equiv\int_{F.C.}\frac{d^4p}{(2\pi)^4}e^{-ipx}\rho(p^2) \ , 
\ee
where $F.C.$ means that the states in a healthy theory will have momenta in the forward lightcone: $p^2>0, p^0>0$. The $\rho(p^2)$ is the spectral density for this 2pt function; in a unitary theory it will be positive.

Now let's consider some examples. For $\Delta_\phi=1$,  it is easy to show that in momentum space the spectral density corresponds to the one of free massless scalar particles (concentrated on the lightcone):
\be{delta1}
\frac{1}{x^2}\rightarrow \rho(p^2)=\delta(p^2) \ ,
\ee
For $\Delta_\phi=2$, we see that
\be{delta2}
\frac{1}{x^4}\rightarrow \rho(p^2)= p^2\propto \int_{F.C.} d^4q\, \delta[(p-q)^2]\delta(q^2) \ .
\ee
This is not concentrated on the lightcone, but is given by the phase space of two massless particle states.

For $\Delta_\phi=1+\gamma$ non integer, 
\be{deltanon}
\frac{1}{(x^2)^{1+\gamma}}\rightarrow \rho(p^2)\sim(p^2)^{\gamma-1} \ .
\ee
We see that the spectrum of theory, the set of $p^2$ values at which $\rho(p^2)\ne0$, is continuous: it spreads from $0$ to $+\infty$. Also there is no way to represent it by a phase space of finitely many massless particles, like for the $\phi^2$ operator.
Naively we have a fractional number of particles. More correctly, such a theory simply cannot be interpreted in terms of particles. In such a theory there is no S-matrix and the only observables are the correlation functions.

Another Type C example, with an important phenomenological meaning, is found in the $\lambda\phi^4$ theory in $2\le D< 4$ dimensions:
\be{phi4}
\mathscr L =\frac{1}{2}(\partial\phi)^2+m^2\phi^2+\lambda\phi^4 \ , 
\ee
In $D<4$, both $\phi^2$ and $\phi^4$ are relevant operators,\footnote{Recall that relevant (irrelevant) operators are those of dimension $\Delta<D$ ($\Delta>D$). Operators of dimension $\Delta=D$ are called marginal.}
 and their couplings have positive mass dimension. Let us express them in terms of the renormalization scale $\mu$:
\be{redefml}
m^2=t\mu^2 \ \ \ \text{and} \ \ \ \lambda=\bar\lambda\mu^{4-D} \ ,
\ee
with $t,\bar\lambda $ dimensionless. Let us analyze the structure of the above theory by using the beta functions
\begin{equation}
\begin{aligned}
\label{betaphi4}
\beta(\bar\lambda)&=-(4-D)\bar\lambda+c_1t^2+c_2\bar\lambda^2+\ldots\\
\beta(t)&=-2t+c_3t^2+c_4t\bar\lambda+c_5\bar\lambda^2+\ldots \ , 
\end{aligned}
\end{equation}
with $c_i$ various constants.\footnote{Some of these constants are regulator-dependent. E.g.~in dimensional regularization we have $c_1=c_3=c_5=0$.}

We start the RG flow in the UV with $t,\bar\lambda$ very close to zero. We see that both couplings start growing, since they are both relevant. In $D=4$, the beta function of $\bar\lambda$ does not have a fixed point at one-loop order. In $D=4-\epsilon$ there is a fixed point $(\lambda_*,t_*)$ where both beta functions vanish:
\be{W-F}
\begin{aligned}
\beta(\bar\lambda_*)=\beta(t_*)=0\ , \bar\lambda_*=O(\epsilon), t_*= O(\epsilon^2) \ .
\end{aligned}
\ee
This is called the Wilson-Fisher fixed point. Perturbative analysis can be trusted only for $\eps\ll1$, but the fixed point actually exists from $D=4-\epsilon$ all the way down to $D=2$.

At the IR fixed point the $\phi^4$ will become irrelevant, but the $\phi^2$ operator will still be relevant. This fact has important consequences for the physics of this theory, as can be seen in the RG flow diagram:
\begin{center}
\includegraphics[scale=0.7]{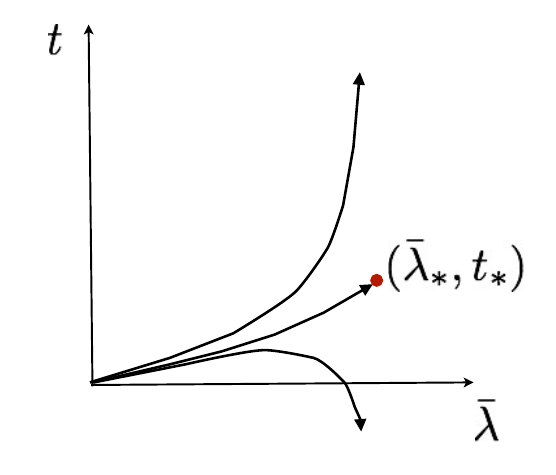}
\end{center}

There is one RG trajectory connecting the origin (free theory) to $(\bar\lambda_*,t_*)$ and one has to choose the initial conditions to lie precisely on this trajectory to reach the fixed point. If the initial value of $t$ is slightly different, then we end up with a theory of massive particles, either $t\to\infty$ so that the $\bZ_2$ symmetry is preserved, or $t\to-\infty$ where the $\bZ_2$ symmetry is broken. The source of this instability is the fact that $\phi^2$ is a relevant operator even at the fixed point.

So we can classify fixed points into two categories:
\begin{itemize}
\item[1.]{Stable --- theories that do not contain relevant scalar operators which are singlets, in the sense explained in the next item. }
\item[2.]{Unstable --- theories in which there exist relevant scalar operators that are singlets under all global (internal) symmetries. In these theories, the coefficients of these operators cannot be naturally assumed to be small, so fine-tuning is necessary.  }
\end{itemize}
The hierarchy problem of the Standard Model can be seen as the fact that the free scalar theory in $D=4$ is an unstable scale invariant theory, the operator $|H|^2$ being a relevant operator and singlet under all global symmetries.

The Banks-Zaks fixed point is stable. The operator $\bar\psi\psi\equiv\bar\psi_L\psi_R+h.c.$ is relevant, but it's not a singlet under the global symmetry $SU(N_f)_L\times SU(N_f)_R$.

Consider next a theory from statistical mechanics, the Ising model, which is a microscopic model for ferromagnetism. Its Hamiltonian (i.e. Euclidean action) is given by
\be{isham}
H=\frac{1}{T}\sum_{<ij>}(1-s_i s_j) \ ,
\ee 
where $T$ is the temperature. The summation is understood to be over neighboring points of a cubic lattice in $\mathbb{R}^D$ and the spins can take the values $s=\pm 1$.

\begin{figure}[!h]
\centering
\includegraphics[scale=.5]{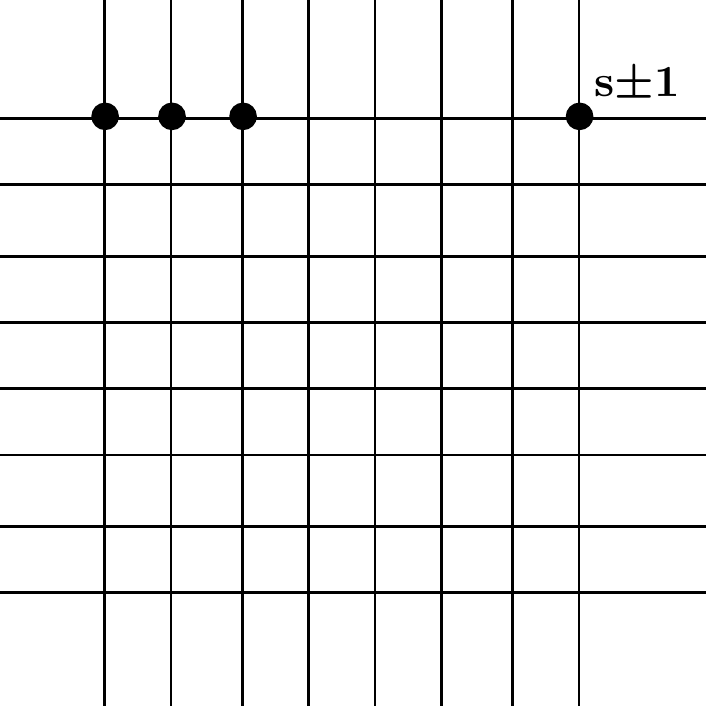}
\end{figure}

The interesting observables are the spin correlation functions. The 2pt function behaves at large distances as
\be{iscorr1}
\langle s(r)s(0)\rangle \sim e^{-r/\xi(T)} \ ,
\ee
where $\xi(T)$ is the correlation length. This correlation length is the analogue of inverse mass in particle physics. Indeed, consider the free scalar propagator in Euclidean coordinate space:
\be{propagator}
\int \frac{d^Dp}{(2\pi)^4}\frac{e^{ipx}}{p^2+m^2}\sim\frac{e^{-rm}}{r^{D-2}} \ . 
\ee
Large distances are $r\gg 1/m$ and we can identify $\xi(T)=1/m$. As is well known, this model has a critical temperature $T=T_c$, at which $\xi(T_c)=\infty$. This critical theory is scale invariant with continuous spectrum (Type C). In fact the critical theory is identical to the Wilson-Fischer fixed point in $D$ dimensions. This is due to a phenomenon called \textit{universality}: that in the continuum limit microscopic details of the Lagrangian don't matter and all theories with the same symmetry look the same (up to identification of couplings).

In the Ising model, to reach the critical point we have to fine-tune the temperature to its critical value $T_c$. This finetuning is the same as the one needed in to reach the Wilson-Fisher fixed point in the $\lambda\phi^4$ theory.

\section{Existing techniques}

Our goal will be to develop methods which allow us to ``solve" fixed points. By "solve" we mean compute the observables: the operator dimensions and the correlation functions. The existing methods include:
\begin{itemize}
\item[1.]{Monte-Carlo simulations for lattice models. This is an established method, whose advantages are that it's fully non-perturbative, from first principles, and does not involve theory biases or unproven assumptions. The downsides is that the computer cost may be significant (especially in higher dimensions), and that passing to the continuum and to the infinite volume limits requires extrapolation.}
\item[2.]{High-temperature expansion for lattice models. Consider again the Ising model in \eqref{isham} with partition function $Z=\sum\exp[-H]$ and expand the exponential for large $T$. The coefficients of the resulting power series in $1/T$ can be evaluated explicitly for the few (a couple of dozens before the combinatorics becomes overwhelming) terms. This is the same as the strong coupling expansion in field theory. The series will start diverging for $T$ near $T_c$ and extrapolating into this region one gets information about the critical point.}

\item[3.]{The $\epsilon$-expansion. If we work in $D=4-\epsilon$, the loop expansion coincides with the expansion in $\epsilon$ and the critical point is weakly coupled. Setting $\epsilon=1$, one can try to recover the physical case of 3d ferromagnets. However the series obtained this way are divergent starting from the second or third term. In order to get stable results, one has to resum these series using procedures like the Borel resummation etc. }
\end{itemize}
Each of the above techniques has different sources of systematic errors. Results from the different techniques largely agree among themselves (and with the experiment), giving us confidence that they are correct.
\begin{itemize}
\item[4.]{The Exact Renormalization Group (ERG)}. 
\end{itemize}

Consider a lattice theory and do a block-spin transformation, i.e. unite spins into blocks: 
\begin{center}
\includegraphics[scale=1]{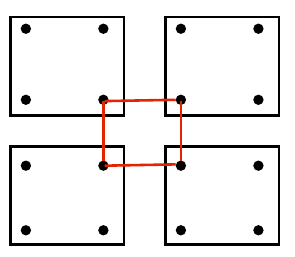}
\end{center}

The partition function remains the same in terms of the new block variables, but the Hamiltonian is changed since there will appear diagonal, quartic couplings etc. (We assume that the original Hamiltonian had only nearest neighbor couplings) 

Repeating this procedure several times, we end up with a new Hamiltonian with different interaction terms. The ERG is a flow in this infinite dimensional space of Hamiltonians. The fixed point is defined as the Hamiltonian which is invariant under this transformation:
\be{ERG}
\mathcal R[H]=H \ .
\ee
Suppose we have found such a Hamiltonian. The next step is to extract scaling dimensions of the local operators. Let's look for perturbations of the fixed point Hamiltonian:
\be{loc1}
H+\Delta H \ ,
\ee
which transform homogeneously under the renormalization group transformation:
\be{loc3}
\mathcal R[H+\Delta H]=H+\lambda \Delta H\,.
\ee
I.e. we are solving an eigenvalue problem for the ERG transformation linearized around the fixed point.
The eigenvector perturbations will be in one-to-one correspondence with the local operators $\calO_\Delta$ at the fixed point:
\beq
\Delta H\sim \int d^Dx\,\calO_\Delta(x)\,,
\eeq
while the eigenvalue 
$\lambda$ will be related to the operator dimension: $\lambda=2^{D-\Delta}$ (2 because we are changing scale by a factor of 2).

The fixed point local operators of definite scaling dimension, $\calO_\Delta$, would look very complicated if expressed in terms of the original lattice variables. In principle, they would be given by a sum of infinitely many terms allowed by the symmetry. E.g. the spin field operator $\sigma$ with definite scaling dimension $\Delta_\sigma$ will be given by something like
\beq
s_i+\text{3 spin terms}+\text{5 spin terms}+\ldots
\eeq
while the first $\mathbb{Z}_2$ even operator will be given by
\be{loc2}
s_is_{i+1}+\text{2 spin terms at larger separation}+ \text{4 spin terms}+\ldots \ ,
\ee
etc.

When analyzing the ERG equation \reef{ERG} in practice, one has to truncate. For example one can allow all couplings within some large distance $N$. One famous example, for the 2d Ising model, is in Wilson's 1975 Reviews of Modern Physics \cite{Wilson:1974mb}. Computing the fixed point Hamiltonian in the space of 217 couplings he found the scaling dimensions in good agreement with the exact solution. Also, he saw that the off-diagonal, quartic, etc. couplings are suppressed, e.g. the long distance couplings $s_i s_{i+n}$ were rapidly decreasing with $n$,
so that the fixed-point Hamiltonian remains relatively short ranged.

There exists also a continuous version of the ERG. E.g.~start with a scalar theory and add all operators 
\be{scalar1}
(\partial\phi)^2+\sum_i c_i \Lambda^{D-\Delta}\mathcal O_i \ , 
\ee
with, in principle, infinitely many couplings $c_i$. This only makes sense in presence of a momentum cutoff, which can be chosen rotationally invariant and smooth. RG flow is performed by integrating out an infinitesimal momentum shell. Also in this case one is forced to truncate. One reasonable truncation is to include all powers of $\phi$ but not the terms with derivatives. This ``local potential approximation" can be motivated by the hope that the fixed point Hamiltonian remains rather local, so that the expansion in derivatives may work.

Limitations of this technique become apparent if we consider a gauge theory. All gauge-invariant operators include more derivatives and more powers of the gauge field simultaneously, so there is no meaningful truncation. The situation is somewhat better in presence of fermions, as here we can privilege arbitrary powers of $\overline\psi\psi$. So the QCD conformal window might be amenable to the ERG. Another bad case is gravity, since all higher dimensional operators like $R^2, R_{\mu\nu}R^{\mu\nu}, \ldots,$ contain more and more derivatives of the metric. So looking for a UV fixed point of gravity with this technique, as in the ``asymptotic safety" program, is very dubious.\footnote{Another problem with asymptotic safety in \emph{gravity}, as opposed to QFT, is that the very meaning of the fixed point is unclear in this case. In quantum field theory, fixed points can be defined axiomatically through the CFT, as we will see later in these lectures, so the ERG is just one of the many ways to get at something which exists independently of the ERG. In case of gravity the fixed point is, until now, \emph{defined} through the ERG manipulations, and an independent axiomatic definition is unknown. Finding such a definition is very important if one hopes to put asymptotic safety on solid ground.}

In summary, the ERG seems to work when we know that a fixed point exists and that the fixed point Hamiltonian is approximately local. It's not clear how reliable this method is to search for new fixed points, or to describe fixed points which do not allow for a local description in terms of the original variables.

The above tools are all based on the RG idea: define a theory at the microscopic level and study its behavior in the IR. Philosophically, this may seem not very satisfactory. The fixed points are universal mathematical objects. It is not clear why in order to solve them we have to approach them with a flow. Can't we solve the fixed points by defining them through some axioms and studying them in isolation? This is in fact possible, and will be the subject of the rest of these lectures.

\section{Towards a nonperturbative definition}
We need a non-perturbative definition of a fixed point theory which does not make any reference to the microscopic level (Lagrangian etc). 

\subsection{Operator spectrum}
The first thing which characterizes any such theory is \underline{the spectrum of the local operators} 
\be{spec}
\mathcal O_i\rightarrow \Delta_i=\text{scaling dimension}  \ .
\ee
Once we know the dimension, the 2pt function is given by
\be{2pfop}
\langle\mathcal O_i (x)\mathcal O_i(0)\rangle=\frac{c}{\vert x\vert^{2\Delta_i}} \ ,
\ee
where the coefficient $c=1$ can be chosen as a normalization convention. Scale transformations are written as 
\be{ST}
x\rightarrow\lambda x \ , \ \ \ \mathcal O(x)\rightarrow {\mathcal O}(\lambda x)=\lambda^{-\Delta}\mathcal O(x) \ .
\ee
With this definition, the 2pt functions are invariant
\be{2pfopST}
\langle{\mathcal O}(\lambda x_1){\mathcal O}(\lambda x_2)\rangle=\frac{1}{\vert \lambda x_1-\lambda x_2\vert^{2\Delta}}=\lambda^{-2\Delta}\langle\mathcal O(x_1)\mathcal O(x_2)\rangle\ .
\ee
The scale transformation can be understood physically as an RG transformation which leaves the Hamiltonian and the correlation functions invariant, as long as the operators are appropriately rescaled. 

The above discussion concerned scalar operators, but there will be operators with nonzero spin as well. We consider Lorentz (or rotation) invariant theories, so the operators will come in irreducible representations of the $SO(D)$ group.

\subsection{Stress tensor and currents}

Among the local operators of the theory, a special role will be played by the \underline{stress tensor} $T_{\mu\nu}$ and \underline{conserved currents} $J_\mu$ associated to global symmetries. The minimal set of QFT axioms (Wightman axioms) don't require existence of the stress tensor as the energy and momentum density, but only of the full energy and momentum charges, and analogously for the conserved currents. However, the existence of these operators is a natural extra assumption. It means that the theory preserves some locality. 

If the IR fixed point can be reached from a UV theory which has a weakly coupled Lagrangian description (and thus has a stress tensor), then the existence of a stress tensor in the IR is guaranteed. On the other hand, if we reach the critical point starting from a lattice, the stress tensor existence is not obvious. On the lattice there is no stress tensor, but it may emerge, together with the rotation invariance, in the continuum limit. This is what happens if the lattice interactions are local, like in the nearest-neighbor Ising model, but it's not guaranteed.

Let us consider a physical example without a stress tensor. Examples of such kind are of interest for the condensed matter and statistical physics. Start from a lattice theory with an explicitly non-local Hamiltonian
\be{latham}
H=\frac{1}{T}\sum_{i,j}\frac{(1- s_is_j)}{\vert i-j\vert^{D+\sigma}}\ ,
\ee
where the sum is over all pairs of spins (not just neighboring ones). This is called the ``long-range Ising model", and it is known to have a critical point in $2\le D<4$ for any value of $\sigma>0$. The properties of this critical point depend on $\sigma$.
For large $\sigma$,  $\sigma>\sigma_*$, the critical point is identical to that of the usual, short-range Ising model (and thus has a local stress tensor). This is because for large $\sigma$ the function $1/{\vert i-j\vert^{D+\sigma}}$ is highly peaked at $i=j$. So we are in the same universality class as the usual Ising model.

On the other hand, for small $\sigma<\sigma_0$, the resulting theory is simply Gaussian (non-interacting), so that it is completely characterized by the 2pt function
\be{gauss}
\langle s(r)s(0)\rangle \sim\frac{1}{r^{2\Delta}}  \ , 
\ee 
from which the higher-point functions are computed by Wick's theorem. The dimension $\Delta$ depends on $\sigma$. This theory does not have a local stress tensor, i.e.~a spin-2 local conserved operator. It can also be described as the AdS dual of a free scalar theory in the bulk with gravity turned off, which also explains that it does not have a local stress tensor.

In the intermediate range, $\sigma_*<\sigma<\sigma_0$ the critical point is an interacting (non-Gaussian) theory, which again does not have a local stress tensor. 

Below by stress tensor we will always mean a local stress tensor and will drop `local'.

Let's go back to the more familiar case of theories which do have a stress tensor and currents. Notice that in perturbation theory these two operators don't renormalize, i.e.~they have zero anomalous dimensions: $\gamma_T=\gamma_J=0$. This means that at the fixed point the stress tensor and currents have canonical dimensions, i.e. 
\be{tjdims}
\Delta_T=D\ , \Delta_J=D-1 \ .
\ee
They are also conserved:
\be{conserv}
\partial_\mu T^{\mu\nu}=0 \ \ \ \text{and} \ \ \ \partial_\mu J^{\mu}=0 \ .
\ee

\subsubsection{Ward Identities}

Consider the Ward identities for an operator $\mathcal O_1$ in the UV. The Ward identity is usually written in the differential form:
\be{WI}
\partial_\mu\langle T^{\mu\nu}(x)\mathcal O_1(x_1)\ldots\rangle=\delta(x-x_1)\langle \partial_\mu\mathcal O_1(x_1)\ldots\rangle \ .
\ee
Since this involves the $\delta$-function, it may seem that the Ward identity is sensitive to the UV form of correlators. However, when we flow to the IR we expect that the IR Ward identity should be valid in a form which makes no reference to the UV. To see that, start from the integrated Ward identities
\be{IWI}
\int\partial_\mu a(x)\langle T^{\mu\nu}(x)\mathcal O_1(x_1)\ldots\rangle=a(x_1)\langle \partial_\mu\mathcal O_1(x_1)\ldots\rangle \ ,
\ee
and choose $a(x)$ to be equal to 1 (0) inside (outside) a sphere centered at $x_1$. Therefore,
\be{IWI2}
\int_{S^{D-1}}d\Sigma \ n_\mu\langle T^{\mu\nu}(x)\mathcal O_1(x_1)\ldots\rangle=\langle \partial_\mu\mathcal O_1(x_1)\ldots\rangle \ ,
\ee
where $S^{D-1}$ is the sphere in $D$ dimensions. We have the liberty to choose the radius of the sphere to be arbitrarily large, therefore the above relation will hold in the IR. 

The moral of this story is that the IR limit of the correlators forms a consistent local theory by itself. There is no need to keep the UV tails of the correlators - we can cut them off and replace by the IR correlators extrapolated to short distances.

\subsubsection{The trace of the stress tensor}

$T_{\mu\nu}$ can be decomposed into two parts transforming irreducibly under $SO(D)$, the trace and the symmetric traceless part. Generically different $SO(D)$ representations are expected to have different scaling dimensions. But the trace of the stress tensor has the same dimension as the stress tensor itself: $[T^\mu_{ \ \mu}]=D$. The resolution of this minor paradox is that in fact, generically, $T^\mu_{\ \mu}=0$ at fixed points. 

In perturbation theory, this can be seen as a consequence of 
\be{perttheory}
T^\mu_{\ \mu}\sim\beta(g_i)\mathcal O_i \ ,
 \ee
 where $\calO_i$ is the operator multiplying the coupling $g_i$ in the Lagrangian. This equation has subtleties associated with it, especially in the multi-field case with broken global symmetries. Ignoring these subtleties, we conclude that the trace of the stress tensor is zero at fixed points, where the beta function vanish.  

Another argument, not relying on perturbation theory, is as follows. The RG transformation, which rescales globally the coordinates, can be seen as a rescaling of the metric:
\be{sct}
x\rightarrow x'=\lambda x\ , \ \ \ g_{\mu\nu}\rightarrow g'_{\mu\nu}=\lambda^2 g_{\mu\nu} \ ,
\ee
where $g_{\mu\nu}\equiv\delta_{\mu\nu}$. For $\lambda-1=\eps\ll 1$ the metric change is small: $\delta g_{\mu\nu}=2\eps\delta_{\mu\nu}$. Use a definition of the stress tensor as an operator measuring response to changing the metric, the Hamiltonian changes by
\be{prop}
\Delta H=\int d^Dx\, T_{\mu\nu}\,\delta g^{\mu\nu}\propto \int d^Dx\, T^\mu_{\ \mu} \ .
\ee
This must be zero if the theory is scale invariant. Generically this means that, in a scale invariant theory, the trace of the stress tensor must be a total divergence of a vector operator:
\be{totdiv}
T^\mu_{\ \mu}=\partial_\mu K^\mu \ , \ \ \ [K^\mu]=D-1 \ . 
\ee
We see that $K_\mu$ has canonical dimension. However generically, all vector operators acquire anomalous dimensions apart from those which are conserved. Thus generically $\partial_\mu K^\mu=0$, and so we conclude again that $T^\mu_{\ \mu}$ must vanish.

One should stress however that rare examples of scale invariant theories with $T^\mu_{\ \mu}\ne0$ do exist. An example is the theory of elasticity in the Euclidean space
\be{elast}
\mathscr L = a (\partial_\nu u_\mu)^2+b (\partial_\mu u_\nu)^2 \ , 
\ee
with $a \ \text{and} \ b$ the elasticity moduli. Physical cases are $D=2,3$. This theory, although interesting and physically motivated, is not unitary. More precisely, it lacks a property called reflection positivity, which is the Euclidean analogue of unitarity and which we will discuss in later lectures. The exception is $a=-b$, in which case we get a gauge invariant Maxwell theory which \emph{is} reflection positive (and unitary upon continuation to the Minkowski space). Only in this case and in $D=4$ the theory has $T^\mu_{\ \mu}=0$.

Above we argued from genericity. In fact, for unitary theories in 2d and 4d (with some extra assumptions in 4d), there are theorems that $T^\mu_{\ \mu}=0$.

\subsection{Conformal invariance}

From now on we assume that $T^\mu_{\ \mu}=0$. As we will see now, this implies that SI of a theory is enhanced to conformal invariance (CI). Indeed, the Hamiltonian is now invariant under metric deformations $\delta g_{\mu\nu}=c(x)\delta_{\mu\nu}$, where $c(x)$ can be an arbitrary function of coordinates:
\be{prop1}
\Delta H=\int d^Dx\, T_{\mu\nu}\,\delta g^{\mu\nu}=\int d^Dx\, c(x)T^\mu_{\ \mu}=0 \ ,
\ee
Transformations changing the metric in this way are also called Weyl transformations. So we are saying that the theory with vanishing $T^{\mu}{}_\mu$ is invariant under infinitesimal Weyl transformations. Invariance under finite Weyl transformations is less obvious, because $T^{\mu}{}_\mu$ will acquire a nonzero vev in a curved even-dimensional spacetimes (the so called Weyl anomaly). Below we only need infinitesimal Weyl transformations and so the Weyl anomaly will not play a role.

A general Weyl transformation (even an infinitesimal one) changes the spacetime geometry from flat to curved, which we don't want to do.
There is however a subclass of infinitesimal Weyl transformations for which the spacetime remains flat. These are those Weyl transformations which can be realized by applying an infinitesimal coordinate transformation (diffeomorphism):
\be{sct2}
x'^\mu= x^\mu+\eps^\mu(x), \ee
and rewriting the metric in new coordinates, which corresponds to:
\be{sct3}
\delta g_{\mu\nu}= \partial_\mu \epsilon_\nu +\partial_\nu\epsilon_\mu\stackrel{\text{must be}}{=} c(x)\delta_{\mu\nu} \ ,
\ee
The last condition is necessary to have a Weyl transformation. Transformations which change the metric this way are called \underline{conformal}.
It turns out that this last equation allows only four classes of solutions in $D\ge 3$ dimensions:
\be{sols}
\begin{aligned}
&\epsilon^\mu=\text{constant}&& \text{infinitesimal translation, } c(x)=0 \ ,\\
&\epsilon^\mu=x^\nu  \omega_{[\nu\mu]}&& \text{infinitesimal rotation, }c(x)=0 \ ,\\
&\epsilon^\mu=\lambda x^\mu &&\text{scale transformation, } c(x)=2\lambda \ ,\\
&\epsilon^\mu=2(a\cdot x)x^\mu-x^2 a^\mu&&\text{Special Conformal Transformations (SCT), } c(x)=a\cdot x, \\
&\text{with} \ a^\mu\text{ an arbitrary vector}.
\end{aligned}
\ee
In $D=2$ dimensions, there are more solutions. To see that, we introduce the complex variable 
\be{zeta}
z=x_1+ix_2,\ \bar z=x_1-ix_2 \ , \qquad ds^2=dzd\bar z \ .
\ee
If we transform $z\rightarrow f(z), \ \bar z\rightarrow f(\bar z)$, with $f$ an analytic function, this corresponds to a conformal rescaling of the metric
\be{transz}
ds^2=\vert f(z)\vert^2 ds^2 \ .
\ee
The solutions for $\delta z$ are
\be{solsz}
\begin{aligned}
&\delta z= \text{constant $\rightarrow$ translation, }\\
&\delta z=e^{i\theta}z \ \text{$\rightarrow$ rotation, }\\
&\delta z=\lambda z \ \text{$\rightarrow$ scale transformation, }\\
&\delta z=c z^2 \ \text{$\rightarrow$ SCT .}
\end{aligned}
\ee
Higher powers $\delta z\propto z^n$, $n>2$, corresponds to new conformal transformations which exist only for $D=2$. 

In these lectures we will mostly talk about $D\ge 3$. Let us integrate the infinitesimal forms of the conformal transformations to find their finite form. For the SCT, which is the only nontrivial case, we find
\be{finiteSCT}
x'^\mu=\beta(x)(x^\mu-a^\mu x^2) , \
\ee
with
\be{betax}
\beta(x)=\frac{1}{1-2(a\cdot x)+a^2x^2} \ .
\ee

A general conformal transformation $x\rightarrow x' $ will be a composition of translations, rotations, scale transformations and SCTs.  The group of conformal transformations is a finite-dimensional subgroup of the group {\it Diff} of diffeomorphisms of $\mathbb{R}^D$. It is in fact the largest finite-dimensional subgroup of {\it Diff}. 

The defining property of conformal transformations, i.e.
\be{GeneralCT}
g'_{\mu\nu}(x')=c(x)\delta_{\mu\nu} \ ,
\ee
means that the Jacobian of the coordinate transformation must be proportional to an orthogonal matrix $M\in SO(D)$:
\be{jacob}
J=\frac{\partial x'^\mu}{\partial x^\nu}=b(x)M^\mu_{\ \nu}(x)\,,\ b(x)=\sqrt{c(x)}\,.
\ee
I.e.~conformal transformations locally looks like a rotation and a scale transformation. 

This allows us physically to think of a conformal transformation as a non-uniform RG transformation. I.e.~we start with a uniform lattice which is mapped by a conformal transformation to a nonuniform but still locally orthogonal one. Then we do a block transformation by uniting spins within new cells.
\begin{center}
\includegraphics[width=5cm]{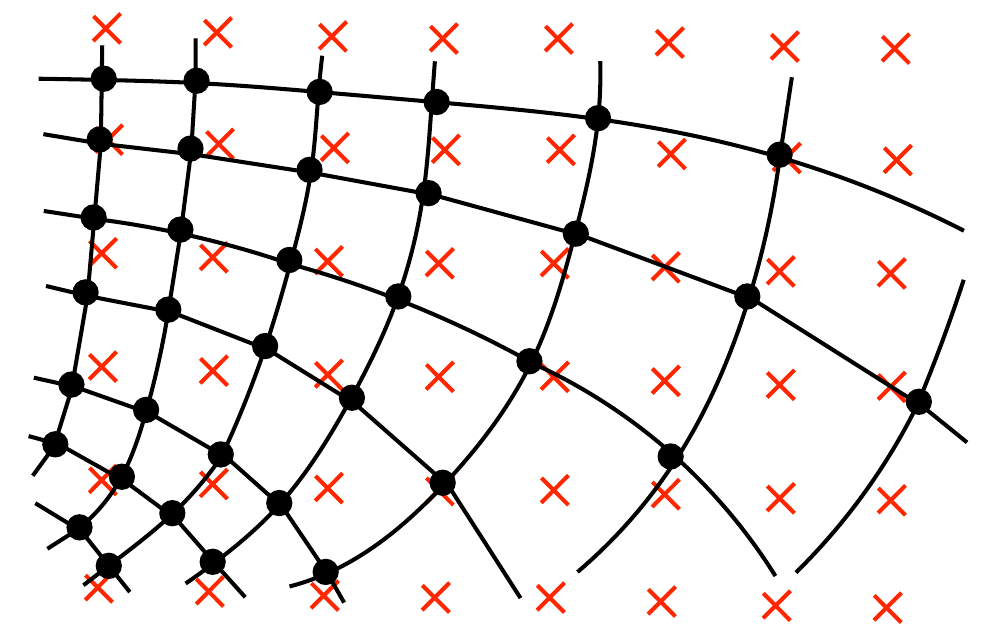}(figure adapted from \cite{Cardy:1987cs})
\end{center}

This way of thinking gives another handle on the question why SI implies CI. It simply means that a system invariant under RG transformations will also be invariant under non-uniform RG transformations. Schaefer in 1976 \cite{Schafer:1976ss} used the ERG and, under the assumption that the fixed point Hamiltonian is sufficiently local, showed that it is also invariant under non-uniform ERG transformations expressing conformal invariance. 

Schaefer's treatment, as the above genericity arguments, illustrate the crucial role of locality in the connection between the SI and CI. The role of unitarity is not as clear. Although natural from the point of view of particle physics, unitarity is an unnecessary and restrictive assumption in statistical mechanics, where many interesting non-unitary fixed points do exist. Empirically, the absolute majority of these fixed points is conformal, and so conformal invariance must have an explanation not involving unitarity.

\subsection{Transformation rule for operators}

Invariance of the Hamiltonian is not the full story. Conformal transformations move points around, and correlation functions with insertions at new and old positions will be related. To understand this rule the non-uniform RG picture introduced above is somewhat useful. We saw earlier that under scale transformations the operators transform as
\be{ST2}
x\rightarrow\lambda x \ , \ \ \ \mathcal O(x)\rightarrow \widetilde{\mathcal {O}}(\lambda x)=\lambda^{-\Delta}\mathcal O(x) \ .
\ee
We should think of $\widetilde{\mathcal{O}}$ as the same operator as $\mathcal{O}$ but in the theory with the RG-transformed Hamiltonian. Since the Hamiltonian is RG-invariant, we can identify the correlators of $\widetilde \calO$ and $\calO$. And so, the above transformation rule means that the 2pt function of $\calO$ should satisfy the following covariance property:
\be{2pfopST2}
\langle{\mathcal O}(\lambda x_1){\mathcal O}(\lambda x_2)\rangle= \lambda^{-2\Delta}\langle\mathcal O(x_1)\mathcal O(x_2)\rangle\ .
\ee
Now, we have to guess how the operators transform under conformal transformations. The simplest assumption is that, if an operator is sufficiently local, it should not feel the effect of the variation of the scale factor $b(x)$ in \reef{jacob}. Then, 
\be{CTlocal}
x \rightarrow x', \ {\phi}(x)\rightarrow\tilde{\phi}(x')={b(x)}^{-\Delta}\phi(x) \ ,
\ee
where again the correlators of $\phi$ and $\tilde\phi$ should be identified.
If this is the case, $\phi(x)$ is called a \underline{primary} operator. Even though all operators transform simply under scale transformations, it is not necessary that they follow the above simple rule under conformal transformations with $b(x)\ne const$. For example, if $\phi$ transforms as a primary, then $\partial_\mu\phi$ does not---it has a homogeneous transformation part, as well as an inhomogeneous part proportional to the derivatives of the scale function. Such derivative operators as $\partial_\mu\phi$ called descendants. As we will see later, all operators in a conformal field theory are either primary or descendants. 

Consider now primary operators with an intrinsic spin. Their transformation rules should depend on the rotation matrix $M^\mu_{\ \nu}(x)$ in \reef{jacob}. An operator in an irreducible representation $R$ of $SO(D)$ will transform as
\be{redSOD}
\phi(x)\rightarrow\tilde \phi(x')={b(x)}^{-\Delta}R[M^\mu_{\ \nu}(x)]\phi(x) \ ,
\ee
where $R[M^\mu_{\ \nu}(x)]$ is a representation matrix acting on the indices of $\phi(x)$.

For example, if $\phi$ is a vector, then $R[M^\mu_{\ \nu}(x)]=M^\mu_{\ \nu}(x)$. So spin-one fields will transform as
\be{curtrans}
\tilde V_\mu(x')=b(x)^{-\Delta}M^\nu_{\ \mu}(x)V_\nu(x) \ .
\ee

Symmetry currents and the stress tensor have spin 1 and 2 respectively, and as we will see later they are primary operators. Generically, there will be infinitely many primaries for each spin.

\section*{Literature}

For the introduction to RG and physical foundations of conformal invariance see the evergreen book by John Cardy \cite{Cardy:1996xt}, as well as his little-known but extremely interesting review \cite{Cardy:1987cs}.

For a ``derivation" of conformal invariance from scale invariance using ERG see Schaefer \cite{Schafer:1976ss}.

For recent work on scale vs conformal invariance see Luty, Polchinski and Rattazzi \cite{Luty:2012ww}, Dymarsky, Komargodski, Schwimmer and Theisen \cite{Dymarsky:2013pqa}, and the review by Nakayama \cite{Nakayama:2013is}.

For the long-range Ising model see \cite{Paulos:2015jfa}.

The theory of elasticity as an example of scale without conformal invariance was discussed in detail (in 2d) by Riva and Cardy \cite{Riva:2005gd}.
  
For the introduction to the global conformal transformations, see chapter 4 of the yellow book \cite{yellow}.

For the ERG example by Wilson, see his famous Reviews of Modern Physics article \cite{Wilson:1974mb}.

\chapter{Conformal kinematics}
\label{lecture2}
\section{Projective null cone}

In lecture \ref{lecture1} we motivated the study of Scale Invariant (SI) fixed points. We also argued that the SI is generically enhanced to Conformal Invariance (CI).
We saw that in $D\ge3$ the group of Conformal Transformations (CT) is finite dimensional and is generated by the Poincar\'e transformations plus dilatations plus Special Conformal Transformations (SCT). 

Then we introduced the concept of primary operators, which transform under CT as
\be{primscalar}
\phi(x)\rightarrow \tilde \phi(x')= \frac{1}{b(x)^\Delta}\phi(x)\ , 
\ee
if $\phi$ is scalar, or 
\be{primspin}
\phi(x)\rightarrow \tilde\phi(x')= \frac{1}{b(x)^\Delta}R[M^\mu_{\ \nu}(x)]\phi(x)\ , 
\ee
if $\phi$ has intrinsic spin, i.e. belongs to an irreducible representation $R$ of $SO(D)$. Here the local scale factor $b(x)$ and the local rotation matrix $M^\mu{}_\nu$ appear in the Jacobian of the conformal transformation, see \reef{jacob}. The correlation functions of $\tilde\phi$ are the same as those of $\phi$, as $\tilde\phi$ can be thought of as an image of $\phi$ under a non-uniform RG transformation leaving the Hamiltonian invariant. Below we will sometimes omit the tilde from $\tilde \phi$.
 
Operationally, the above transformation property simply means that the $n$-point correlation functions of $\phi$ must satisfy
\be{const}
\langle \phi(x')\phi(y')\ldots\rangle=\frac{1}{b(x)^\Delta}\frac{1}{b(y)^\Delta}\ldots\langle \phi(x) \phi(y)\ldots\rangle \ .
\ee
This condition is clearly an important constraint on the correlation functions of the theory, and in this lecture we will study its consequences. There are several ways to work them out, some more pedestrian than others. We will present a method which gives the most information in the least possible time.  

First of all we have to understand better the conformal algebra. Last time we wrote the formulas for the vector fields that correspond to the generators of the group
\be{gener}
\begin{aligned}
&P_\mu=i\partial_\mu \ \rightarrow \ \text{translations,}\\ 
&M_{\mu\nu}=i(x_\mu\partial_\nu-x_\nu\partial_\mu)\ \rightarrow \ \text{rotations,}\\
&D= i x^\mu\partial_\mu \ \rightarrow \ \text{dilatations,}\\
&K_\mu=i(2x_\mu(x^\nu\partial_\nu)-x^2\partial_\mu)\ \rightarrow \ \text{SCTs} \ .
\end{aligned}
\ee
From these expressions for the generators we can compute the commutation relations. Part of it is the Poincar\'e algebra
\be{poinal}
\begin{aligned}
&[M_{\mu\nu},M_{\rho\sigma}]=-i(\delta_{\mu\rho}M_{\nu\sigma}\pm \text{permutations}) \\
&[M_{\mu\nu},P_{\rho}]=i(\delta_{\nu\rho}P_\mu-\delta_{\mu\rho}P_\nu)\ , 
\end{aligned}
\ee
The interesting new relations are
\be{newintr}
\begin{aligned}
&[D,P_\mu]=-i P_\mu\\
&[D,K_\mu]=i K_\mu\\
&[P_\mu,K_\nu]=2i (\delta_{\mu\nu}D-M_{\mu\nu}) \ .
\end{aligned}
\ee
It turns out that this ``conformal algebra" is in fact isomorphic to $SO(D+1,1)$, the algebra of Lorentz transformations in $\mathbb R^{D+1,1}$ Minkowski space. Let us demonstrate this fact.

Consider in the latter space the coordinates
\be{euclcoord}
X^1,\ldots X^D, X^{D+1}, \ X^{D+2}\,,
\ee
where $X^{D+2}$ is the timelike direction. We will also use the lightcone coordinates
\beq 
X^+=X^{D+2}+X^{D+1}, \ X^-=X^{D+2}-X^{D+1} \ . 
\ee
In terms of the above, the mostly plus metric $\eta_{MN}$ in $\mathbb R^{D+1,1}$ is
\be{d+2metric}
ds^2=\sum_{i=1}^{D}(dX^i)^2-dX^+dX^- \ .
\ee
The conformal algebra generators will be identified with the $SO(D+1,1)$ generators as follows
\be{corresp}
\begin{aligned}
J_{\mu\nu}&=M_{\mu\nu}\ , \\
J_{\mu+}&=P_\mu\ , \\ 
J_{\mu-}&=K_\mu\\
J_{+-}&=D\ , 
\end{aligned}
\ee
with $\mu,\nu=1,\ldots, D$. It is understood that $J_{\mu\nu}$ is antisymmetric under $\mu\leftrightarrow\nu$.
Then one can check that the conformal algebra commutation relations coincide with the $SO(D+1,1)$ ones:
\be{claim}
[J_{MN},J_{RS}]=-i (\eta_{MR}J_{NS}\pm \text{permutations})\,.
\ee
One example is
\be{ex1}
[J_{\mu+},J_{\nu-}]\propto \delta_{\mu\nu}J_{+-}+\delta_{+-}J_{\mu\nu} \ .
\ee

\textbf{Exercise:} Check and fix all the constants in the above identification.

This result means that the $D$-dimensional conformal group, while acting in a somewhat non-trivial way on $\mathbb R^D$, acts much more naturally (linearly) on the $\mathbb R^{D+1,1}$ space. In the vector representation we can write 
\be{vecrep}
 X^M\rightarrow\Lambda^M_{\ N} X^N\ , 
\ee
with $\Lambda^M_{\ N}$ an $SO(D+1,1)$ matrix. If we could somehow get an action on $\mathbb R^D$ out of this simple action, then the implications of CI would be easier to understand. To do that, we have to embed the $D$-dimensional space into $\mathbb R^{D+1,1}$ dimensional space, i.e.~to get rid of two extra coordinates.

First of all let's restrict the attention to the null cone in $\mathbb R^{D+1,1}$:
\be{nullcone}
 X^2=0
\ee
This gets rid of one coordinate. Since this constraint is preserved by the action of the group, we don't lose simplicity. 

\begin{figure}[ht!]
\begin{center}
\includegraphics[width=8cm]{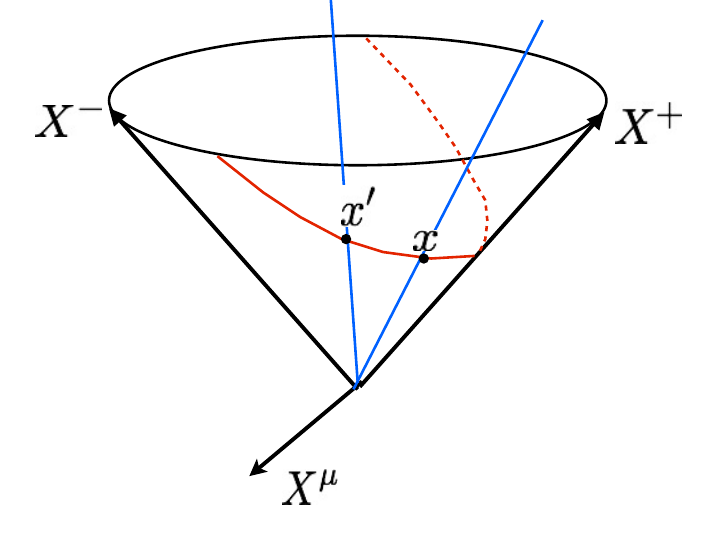}
\caption{Red: section, blue: light ray and light ray$\,'$. }
\label{cone1}
\end{center}
\end{figure}

To get down to $D$ dimensions, we take a generic section of the light-cone:
\be{secti}
X^+=f(X^\mu),
\ee
The section is parametrized by $X^\mu$ which we identify with the $\mathbb{R}^D$ coordinates $x^\mu$

The group $SO(D+1,1)$ acts on the section as follows (see Fig.~\ref{cone1}). A point $x^\mu$ on the section defines a lightray. Applying a Lorentz transformation, this lightray is mapped to a new one which passes through another point $x'^\mu$. Thus
\be{lray}
x^\mu\rightarrow \ \text{light ray} \xrightarrow{\Lambda^M_{\ N}\in SO(D+1,1)} \  \text{light ray}\,' \rightarrow x'^\mu\ .
\ee

\begin{figure}[htbp]
\begin{center}
\includegraphics[width=8cm]{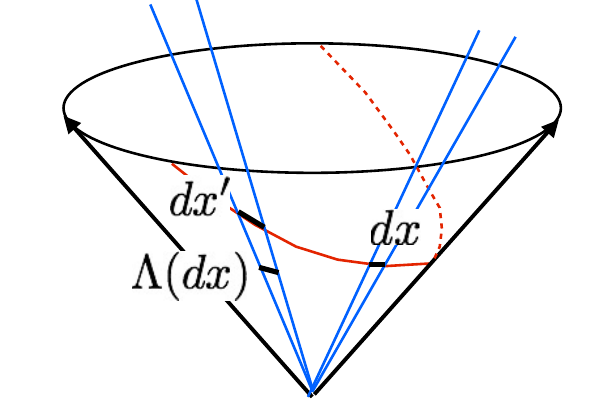}
\caption{How the infinitesimal interval transforms under the defined $SO(D+1,1)$ action. }
\label{cone2}
\end{center}
\end{figure}

This defines an action of $SO(D+1,1)$ on $\bR^D$, and we will now check that this action corresponds to a CT. 
Consider the metric $ds^2$ on the section induced from the Minkowski metric in $\bR^{D+1,1}$. We have:
\be{defindmet}
ds^2=dx^2-dX^+dX^-|_{X^+=f(x), X^-=x^2/X^+}=g_{\mu\nu}(x) dx^\mu dx^\nu\ ,
\ee
where $g_{\mu\nu}(x)$ is a metric we could compute explicitly in terms of $f(x)$ but we won't need it.

The action of $SO(D+1,1)$ on a point $x$ can be split into two steps, 1) $X\to \Lambda.X$ and 2) then rescale to get back into the section. We want to understand how this action changes the infinitesimal interval length (see Fig.~\ref{cone2}) The first step is an isometry and does not change $ds^2$. The second step changes the metric by an $x$-dependent scale factor. Indeed, assuming that we have to rescale by $\lambda$ to get back into the section, where $\lambda$ in general depends on $X$, we have:
\beq
(d(\lambda(X)X))^2=(\lambda dX+X (\nabla \lambda. dX))^2=\lambda^2 dX^2\,,
\eeq
the other terms vanishing by $X^2=0$, $X.dX=0$.

We conclude that the metric transformation is of the form:
\beq
ds'^2=c(x) ds^2,\quad c(x)=\lambda(X)^2\,.\label{resc}
\eeq
This will exactly agree with the definition of the conformal transformations as long as $ds^2$ is flat. From the definition of $ds^2$ it's easy to guess which $f(X)$ achieves this: it is $f(X)=const$, so that $dX^+=0$. For simplicity and without loss of generality we take $const=1$. Thus our \textit{Euclidean} section is parametrized as:
\beq
X^M=(X^+,X^-,X^\mu)=(1,x^2,x^\mu)
\eeq

We note in passing that by taking this section and rescaling it in the radial direction by an $x$-dependent factor 
we can reproduce any metric which is a Weyl transformation of the flat space metric (for example the metric on the sphere, de Sitter or Anti de Sitter spaces).

We would now like to extend the above action to fields. We thus consider fields $\phi(X)$ defined on the cone. The most natural action of the Lorentz group on such scalar fields is
\be{la}
 X\rightarrow  X' \ , \  \phi( X)\rightarrow\tilde \phi( X')=\phi( X)  \ . 
\ee
The field on the Euclidean section will be assumed to coincide with the $D$-dimensional field: 
\beq
\phi( X)|_{\text{section}}=\phi(x)\,.
\eeq
Finally, we will assume that $\phi$ depends homogeneously on $X$:
\be{fieldsec}
\phi(\lambda  X)=\lambda^{-\Delta}\phi( X) \ ,
\ee
Let us show that these conditions imply the correct transformation rule for the fields on $\bR^D$:
\be{fieldgen}
\phi(x')=b(x)^{-\Delta}\phi(x) \ . 
\ee
Indeed, $b(x)$ in this equation is the local expansion factor, and according to Eq.~\reef{resc} it must be identfied with $\lambda(X)$, the scale factor in the second phase of $SO(D+1,1)$ action. Since $\phi(X)$ scales homogeneously with $\lambda$, we get exactly what we need.

Using this ``projective light cone" formalism, any conformally invariant quantity (e.g. correlation function) in $\mathbb R^D$ can be lifted to an $SO(D+1,1)$-invariant quantity in $\mathbb R^{D+1,1}$. Basically, this formalism makes CI kinematics as simple as the Lorentz-invariant kinematics. 

\section{Simple applications}

\subsection{Primary scalar 2pt function}
The expression of the 2pt function on the light-cone is
\be{2pflight}
\langle \phi(X)\phi(Y)\rangle=\frac{c}{(X\cdot Y)^\Delta}\ , 
\ee
with $c$ a constant and $\Delta$ the field's scaling dimension. The above is the most general Lorentz invariant expression consistent with scaling of both $\phi(X)$ and $\phi(Y)$ with degree $\Delta$. Note that $X^2=Y^2=0$ cannot appear. To write the 2pt function in the physical space, we project $X$ and $Y$ on the section, i.e. 
\be{pjojX}
X=(X^+,X^-,X^\mu)=(1,x^2,x^\mu) \  \text{and} \ Y=(Y^+,Y^-,Y^\mu)=(1,y^2,y^\mu) \ .
\ee
We get
\be{XY}
\begin{aligned}
X\cdot Y&=X^\mu Y_\mu-\frac{1}{2}(X^+Y^-+X^-Y^+)\\
&=x^\mu y_\mu-\frac{1}{2}(x^2+y^2)\\
&=-\frac{1}{2}(x-y)^2 \ .
\end{aligned}
\ee
The 2pt function \eqref{2pflight} is therefore projected to
\be{2pfproje}
\langle \phi(x)\phi(y)\rangle\propto\frac{1}{(x-y)^{2\Delta}}
\ee

That this expression is consistent with the SI of the field $\phi(x)$ is obvious. It's less obvious that it's conformally invariant but our construction guarantees it. If we wanted to check this in a pedestrian way, without using the projective light cone, we would have to show that
\be{confequiv}
\vert x'-y'\vert^2= b(x)b(y)\vert x-y\vert^2 \ .
\ee
for any CT, which does not look obvious at first sight. The standard way to derive this is to show first that this holds for the inversion transformation
\be{inverstran}
x'^\mu\rightarrow \frac{x^\mu}{x^2} \ ,
\ee
which is a CT.  Indeed the Jacobian is given by
\be{invchangecoord}
\frac{\partial x'^\mu}{\partial x^\nu}=\frac{1}{x^2}\left(\delta^{\mu\nu}-\frac{2x^\mu x^\nu}{x^2}\right)\equiv b_\text{inv}(x)I^{\mu\nu}(x) \ ,  
\ee
where $b_{inv}(x)=1/x^2$ and $I^{\mu\nu}(x)$ an orthogonal matrix. This can be easily seen if we go to a particular frame where $x$ lies on the $x_1$ direction. Then the matrix is diagonal
\be{Idiag}
I^{\mu\nu}(x)=\begin{pmatrix}
-1&\\
&1\\
& & \ddots\\
& & & 1
\end{pmatrix} \ , 
\ee
and is clearly an $O(D)$ matrix. But it is not in $SO(D)$.
This means that inversion is not in the connected part of the conformal group, i.e. it cannot be obtained by exponentiating a Lie algebra element. 

If we apply inversion twice we get back to the connected component. In fact we can reproduce SCT this way:
\be{SCTa}
\text{SCT}_\mu=\text{inversion}\circ \text{translation}_\mu\circ \text{inversion} \ .
\ee

Going back to Eq. \eqref{confequiv}, it is not difficult to verify for the inversion:
\be{confinvproof}
\vert x'-y'\vert^2=\left\vert\frac{x^\mu}{x^2}-\frac{y^\mu}{y^2}\right\vert^2=\frac{\vert x-y\vert^2}{x^2y^2}=b_\text{inv}(x)b_\text{inv}(y)\vert x-y\vert^2 \ . 
\ee
By \reef{SCTa}, it then holds for SCTs, and by extension for all other CT's. This way of checking the invariance may look a bit ad hoc, and it would become more and more awkward as we go to fields with spin and higher order correlation functions. As we will see, the projective lightcone formalism extends rather easily to such more general situations.

But first two more comments about the 2pt functions of scalar primaries. If the fields have different scaling dimensions, $\Delta_1\neq\Delta_2$, the 2pt function vanishes 
\be{com1}
\langle \phi_1(x)\phi_2(y)\rangle=0 \ .
\ee
This is clear from the lightcone as we cannot construct the analogue of \reef{2pflight} if $\Delta_1\neq\Delta_2$.

In a theory with several fields $\phi_i$ with same scaling dimension $\Delta$, the 2pt function is
\be{com22pf1}
\langle \phi_i(x)\phi_j(y)\rangle=\frac{M_{ij}}{(x-y)^{2\Delta}} \ , 
\ee
As we will see in lecture \ref{lecture3}, the matrix $M_{ij}$ will be positive definite in a unitary theory. This implies  that there exists a field basis such that $M_{ij}$ becomes unit-diagonal: 
\be{com22pf2}
\langle \phi_i(x)\phi_j(y)\rangle=\frac{\delta_{ij}}{(x-y)^{2\Delta}} \ .
\ee
We will always assume that such a basis is chosen.

\subsection{Primary scalar 3pt function}

The 3pt function of three primary scalar fields with scaling dimensions $\Delta_1,\Delta_2,\Delta_3$ (which could be equal or different) must have the following form on the cone
\be{3pf1}
\langle \phi_1(X_1)\phi_2(X_2)\phi_3(X_3)\rangle=\frac{const.}{(X_1 X_2)^{\alpha_{123}}(X_1 X_3)^{\alpha_{132}}(X_2 X_3)^{\alpha_{231}}} \ .
\ee
As in the 2pt function case, the above is the most general Lorentz-invariant expression. To make it consistent with scaling, we should impose the constraints
\be{constr3pf}
\begin{aligned}
&\alpha_{123}+\alpha_{132}=\Delta_1\,,\\
&\alpha_{123}+\alpha_{231}=\Delta_2\,,\\
&\alpha_{132}+\alpha_{231}=\Delta_3 \ .
\end{aligned}
\ee
This linear system of three equations for three unknowns admits a unique solution:
\be{unsol}
\alpha_{ijk}=\frac{\Delta_i+\Delta_j-\Delta_k}{2} \ .
\ee 
Thus the 3pt function is uniquely determined up to a constant. Projecting to the Euclidean section, we find
\be{3pf2}
\begin{aligned}
\langle \phi_1(x_1)\phi_2(x_2)\phi_3(x_3)\rangle=\frac{\lambda_{123}}{|x_{12}|^{2\alpha_{123}}   
|x_{13}|^{2\alpha_{132}}|x_{23}|^{2\alpha_{231}}} \ , 
\end{aligned}
\ee
where $\lambda_{123}$ a free parameter (a CFT analogue of a ``coupling constant''), and
\be{redx}
x_{ij}=x_i-x_j \ .
\ee
This remarkable formula derived by Polyakov in 1970 gave birth to CFT. To understand its significance, we should compare it with infinitely many functional forms which would be allowed if we imposed only SI:
\be{3pfSI}
\sum\frac{const.}{\vert x_{12}\vert^a\vert x_{13}\vert^b\vert x_{23}\vert^c}\ , \ a+b+c=\Delta_1+\Delta_2+\Delta_3 \ ,  
\ee
whereas there is only one term consistent with CI. 

The 3pt function $\langle\sigma(x)\sigma(y)\epsilon(z)\rangle$ for two spins and energy in the 2-dimensional Ising model at the critical point can be extracted from the exact Onsager's lattice solution. Polyakov noticed that it agrees with his formula and conjectured the CI of the critical 2d Ising model, which as we now know is indeed true. 

\subsection{Four point function}

Moving to the 4pt function, consider four identical fields for simplicity. Requiring consistency under Lorentz transformations and scaling, we get on the cone
\be{4pf1}
\langle \phi(X_1)\phi(X_2)\phi(X_3)\phi(X_4)\rangle=\frac{1}{(X_1\cdot X_2)^\Delta(X_3\cdot X_4)^\Delta}f(u,v) \ , 
\ee
Here $u$ and $v$ are \emph{conformally invariant cross-ratios} which on the light-cone are given by Lorentz-invariant expressions
\be{CIu}
u=\frac{(X_1\cdot X_2)(X_3\cdot X_4)}{(X_1\cdot X_3)(X_2\cdot X_4)}\ , \ \ \ \text{and} \  \ \ v=u\vert_{2\leftrightarrow4}=\frac{(X_1\cdot X_4)(X_2\cdot X_3)}{(X_1\cdot X_3)(X_2\cdot X_4)} \ .
\ee
Notice that they have scaling zero in every variable. Since the prefactor in the RHS of \reef{4pf1} takes care of the scaling, any function $f(u,v)$ of $u$ and $v$ can appear.

Projecting to the Euclidean section, $u$ and $v$ become
\be{CIuEs}
u=\frac{x_{12}^2x_{34}^2}{x_{13}^2x_{24}^2} , \ \ \ \text{and} \  \ \ v=u\vert_{2\leftrightarrow4}=\frac{x_{14}^2x_{23}^2}{x_{13}^2x_{24}^2} \ ,
\ee
and the full 4pt function:
\be{4pf3}
\langle \phi(x_1)\phi(x_2)\phi(x_3)\phi(x_4)\rangle=\frac{1}{x_{12}^{2\Delta}x_{34}^{2\Delta}}f(u,v) \ . 
\eeq
That the 4pt function must be given by a simple expression times a function of the conformally invariant cross-ratios is an enormous reduction of the functional freedom, although not as large as for the 3pt functions where the functional form was completely fixed.  

We will later see that $f(u,v)$ is not an independent quantity but is related in a non-trivial way to the 3pt function. This will require a set of arguments going beyond just conformal kinematics.

For the moment let us notice a functional constraint on $f(u,v)$ which comes from the crossing symmetry of the 4pt function. The prefactor of \reef{4pf1} or \reef{4pf3} groups the external points as (12)(34), but this is an arbitrary choice. If we interchange $2\leftrightarrow 4$, we get 
\be{4pf4}
\langle \phi(x_1)\phi(x_2)\phi(x_3)\phi(x_4)\rangle=\frac{1}{x_{14}^{2\Delta}x_{23}^{2\Delta}} f(\tilde u,\tilde v) \ ,
\ee
where now $f(\tilde u,\tilde v)$ depends on the conformally invariant crossratios calculated with the interchanged indices: 
\be{tildeuv}
\tilde u= v\ , \ \ \ \tilde v=u \ .
\ee
Notice that the same function $f$ appears in \reef{4pf3} and \reef{4pf4}, since the 4pt function is totally symmetric under permutations. Moreover, \reef{4pf3} and \reef{4pf4} must agree:
\be{4pf5}
\frac{1}{x_{12}^{2\Delta}x_{34}^{2\Delta}}f(u,v) =\frac{1}{x_{14}^{2\Delta}x_{23}^{2\Delta}} f(v,u) \  
\ee
Multiplying by $x_{14}^{2\Delta}x_{23}^{2\Delta}$. we find that $f(u,v)$ must satisfy:
\be{rat}
\left(\frac{v}{u}\right)^\Delta f(u,v)=f(v,u) \ .
\ee
This constraint will play an important role in lecture \ref{lecture4}.

\section{Fields with spin}

\subsection{Extending the null cone formalism}

So far we only talked about scalar primaries. Let us now consider primaries with spin. 

We will consider symmetric traceless primary fields living on the $D$ dimensional space\footnote{Primaries in other representations of $SO(D)$, like antisymmetric tensors of fermions, can also be considered}. We will put such a field in correspondence with a field which lives on the light-cone and is also symmetric and traceless:
\be{stpf}
\phi_{\mu\nu\lambda\ldots}(x)\leftrightarrow\phi_{MNL\ldots}(X) \ .
\ee
We notice that the fields on the light-cone have more components that the $D$ dimensional ones (roughly two extra components per index). For this correspondence to be useful, we have to eliminate these extra components. Let's first of all impose transversality of the null cone fields
\be{transv}
X^M\phi_{MNL\ldots}(X)=0\ .
\ee
This condition eliminates one extra component per index. We will see below how the remaining ones are dealt with.

Then we define $\phi_{\mu\nu\lambda\ldots}(x)$ to be related to $\phi_{MNL\ldots}(X)$
 by projection on the Euclidean section
\be{projeuclideanphi}
\phi_{\mu\nu\lambda\ldots}(x)=\phi_{MNL\ldots}(X)\frac{\partial X^M}{\partial x^\mu}\frac{\partial X^N}{\partial x^\nu}\ldots \ , 
\ee
where $X^M=(1,x^2,x^\mu)$ is the parametrization of the section, so
\be{jac}
\frac{\partial X^M}{\partial x^\nu}=(0,2x_\nu,\delta^\mu_\nu) \ .
\ee
Notice that this rule preserves the tracelessness condition: if we start from a traceless $(D+2)$-dimensional tensor, we will end up with a traceless $D$-dimensional tensor. Indeed, to compute the trace of $\phi_{\mu\nu\ldots}$ we have to evaluate the contraction
\be{trace1}
\delta^{\mu\nu}\frac{\partial X^M}{\partial x^\mu}\frac{\partial X^N}{\partial x^\nu}
\ee
which can be shown to be equal to 
\beq
\eta^{MN}+X^M K^N+X^N K^M \ ,
\eeq
with $K_M=(0,2,0)$ an auxiliary vector. Contracted with $\phi_{MN\ldots}$, it will vanish by tracelessness and transversality.

Notice also that anything proportional to $X^M$ projects to zero, since
\be{0projec}
X^2=0  \Rightarrow X^M\frac{\partial X^M}{\partial x^\mu}=0 \ .
\ee
This means that, for the purposes of this correspondence, $\phi_{MNL\ldots}$ is defined up to adding an arbitrary tensor proportional to $X^M$. This ``gauge invariance'' further reduces the number of degrees of freedom, exactly to what we need.

Let's discuss the transformation properties. Under an $SO(D+1,1)$ transformation, the field on the null cone transforms in the standard Lorentz-invariant way:
\be{sofield}
\tilde\phi_{MNL\ldots}(X')=\Lambda_M^{\ M'}\Lambda_N^{\ N'}\ldots\phi_{M'N'L'\ldots}(X) \ . 
\ee
Just like for primary scalars, we will impose that the null cone fields are homogeneous in $X$:
\be{scaletransformationscalars}
\phi_{\ldots}(\lambda X)=\lambda^{-\Delta}\phi_{\ldots}(X) \ ,
\ee
We claim that the resulting transformations for the fields on the section is what we need:
\be{conveq}
\tilde\phi_{\mu\ldots}(x')=\frac{1}{b(x)^{\Delta}} M^{\mu'}_{\ \mu}(x)\cdots\phi_{\mu'\ldots}(x) \ .
\ee
The line element transforms as
\beq
dx'=b(x)  M(x).dx
\eeq
To show that \reef{conveq} is true, it's enough to show that (consider spin 1 case for simplicity)
\beq
\tilde \phi(x').dx'=\frac 1{b(x)^{\Delta-1}} \phi(x).dx
\label{tosh}
\eeq
Now, the projection rule implies that
\beq
\phi(x).dx=\phi(X).dX
\eeq
When $X\to \Lambda.X$ the scalar product $\phi(X).dX$ is preserved:
\beq
\phi(Y).dY=\phi(X).dX,\qquad Y=\Lambda.X
\eeq
To get from $Y$ back into the section we have to rescale: $X'=b Y$. When we do it $\phi(Y)$ simply rescales. $dY$ rescales plus gets a contribution proportional to $Y$ if $b$ is not a constant. This extra contribution vanishes when contracted with $\phi(Y)$ because of transversality. The end result is exactly \reef{tosh}

\subsection{Two point function}

To see the consequences, consider the 2pt function of a vector field. On the cone we have: 
\be{2pfv}
\langle \phi_M(X)\phi_N(Y)\rangle=\frac{\eta_{MN}+\alpha\frac{Y_MX_N}{XY}}{(XY)^\Delta}\ ,
\ee
where we once again considered the most general Lorentz invariant form consistent with scaling. Notice that we don't write terms proportional to $X_M$ or $Y_N$, since they anyway project to zero.

We have to impose transversality which fixes the value of the constant $\alpha$
\be{imptransv}
X^M( \ \ \ \ \ )= Y^N ( \ \ \ \ \ )=0 \Rightarrow \alpha=-1 \ .
\ee
Projecting the 2pt function in the physical space, we find:
\be{projvector}
\begin{aligned}
&\eta_{MN}\rightarrow\delta_{\mu\nu}\\
&Y_M\rightarrow -x_\mu+y_\mu\\
&X_N\rightarrow x_\nu-y_\nu\\
&X\cdot Y\rightarrow-\frac{1}{2}(x-y)^2 \ ,
\end{aligned}
\ee
therefore 
\be{fin}
\frac{\eta_{MN}-\frac{Y_MX_N}{XY}}{(XY)^\Delta}\rightarrow \frac{I_{\mu\nu}(x-y)}{(x-y)^{2\Delta}} \ ,
\ee
with
\be{imn}
I_{\mu\nu}(x)=\delta_{\mu\nu}-\frac{2x_\mu x_\nu}{x^2} \ . 
\ee
CI fixed the relative coefficient $-2$ between the two terms here. 
If we had SI only, the relative coefficient would be free.

[It would not be so easy to check that the found 2pt function transforms correctly under CT without using the cone. As usual, it would be sufficient to check that it transforms correctly under inversion. This in turn would be equivalent to the identify
\be{conscheckI}
I(x)I(x-y)I(y)=I(x'-y') \ ,
\ee
with $x'=x/x^2$. One can check this by an explicit computation, expanding throughout, but done this way it looks rather accidental.]

The 2pt function for higher spin primaries can be computed similarly. Interestingly, apart from $I_{\mu\nu}$ , no new conformally covariant tensors appear. All the 2pt functions are made of $I_{\mu\nu}$'s connecting different points, and $\delta_{\mu\nu}$'s if the indices $\mu,\nu$ are associated with the same point.
For example, the 2pt function for a symmetric traceless field will be 
\be{spin22pf}
\langle\phi_{\mu\nu}(x)\phi_{\lambda\sigma}(y)\rangle=\frac{1}{\vert x-y\vert^{2\Delta}}[I_{\mu\lambda}(x-y)I_{\nu\sigma}(x-y)+(\mu\leftrightarrow\nu)+\beta\delta_{\mu\nu}\delta_{\lambda\sigma}] \ ,
\ee
where the term with the delta functions (transforming correctly under CT since both indices get multiplied by the same $M^\mu_\nu$) was inserted in order to satisfy the tracelessness condition, which fixes the value of $\alpha$:
\be{coeal}
\beta=-\frac{2}{D} \ .
\ee

To summarize, the 2pt functions are completely fixed for higher spin primaries just like for the scalar.

\subsection{Remark about inversion}

In ``pedestrian" CFT calculations, not based on the projective null cone formalism, one often checks invariance under the inversion rather than under SCT. However, as we mentioned, inversion is not in the connected part of the conformal group. One may wonder if assuming invariance under the inversion is in fact an extra assumption.

On the cone, the inversion corresponds to the transformation
\be{invlc}
\begin{gathered}
X^{D+1}\rightarrow -X^{D+1},\text{ i.e. }X^+\leftrightarrow X^-\quad(X^{\pm}=X^{D+2}\pm X^{D+1})\,.
\end{gathered}
\ee
Indeed, this maps the point $X^M=(1,x^2,x^\mu)$ on the Euclidean section to
\be{eu}
(1,x^2,x^\mu)\rightarrow(x^2,1,x^\mu)\xrightarrow{\text{rescale}}(1,1/x^2,x^\mu/x^2) \ , 
\ee
which contains inversion in the last component.

Notice that the transformation \reef{invlc} belongs to $O(D+1,1)$ but not to $SO(D+1,1)$, i.e.~it is not in the connected component.

Another transformation in the same class is a simple spatial reflection (parity transformation)
\be{partrans}
X^1\rightarrow -X^1 \ .
\ee
The two discrete symmetries, parity and inversion, are conjugate by $SO(D+1,1)$. 
This implies that a CFT invariant under parity will be invariant under inversion and vice versa. 

There are CFTs which break parity (and inversion). Correlators in those theories, lifted to the null cone, will involve the $(D+2)$-dimensional $\epsilon$-tensor, or $\Gamma_{D+3}$ (the analog of the $\gamma_5$ matrix) for fermions if $D+2$ is even. Since we only considered scalars and symmetric tensors, these structures did not occur in our calculations. 

\subsection{Remark on conservation}

We have seen that for spin-1 and spin-2 primary fields, the form of the 2pt correlation functions is fixed by CI in terms of just one parameter: the scaling dimension of the field. Canonical dimensions
\be{candim}
\begin{aligned}
&\Delta=D-1 \ , \ \ \ \text{for} \ \ \  l=1 \ ,
&\Delta=D \ , \ \ \ \text{for} \ \ \ l=2 \ .
\end{aligned}
\ee
would correspond to the conserved currents and the stress tensor. We expect their 2pt functions to be conserved objects. This should happen automatically since there is nothing to be adjusted. And indeed one can check that this is true. E.g. for the currents
\be{currcons}
\partial^\mu\frac{I_{\mu\nu}(x)}{x^{2\Delta}}=0 \ , \ \ \ \text{for} \ \ \ \Delta=D-1 \ , 
\ee

Notice that the null cone formalism is simply a way to compute constraints imposed by CI. For example, current and stress tensor conservation may be more convenient to check in the physical space rather than on the null cone. There is no reason to insist in doing everything on the null cone. The two points of view - null cone and physical space - can be used interchangeably depending what one wants to compute.

\subsection{Scalar-scalar-(spin $l$)}

The last correlator that we will study in this lecture is the 3pt function of two scalars and one spin $l$ operator. Start with spin one. On the null cone we will have 
\be{cor3sl}
\langle \phi_1(X)\phi_2(Y)\phi_{3M}(Z)\rangle=\text{scalar factor}\times\text{(tensor structure)}_M \ ,
\ee
The scalar factor will be the same as for the scalars
\be{analogsca}
\frac{const.}{(XY)^{\alpha_{123}}(YZ)^{{\alpha_{231}}}(XZ)^{{\alpha_{132}}}} \ ,
\ee
where the powers are fixed by the dimensions of the fields in order to get the correct scaling. The tensor structure must then have scaling 0 in all variables, and will also have to be transverse: $Z^M(\ldots)_M=0$. Moreover we don't need to include a term proportional to $Z_M$ since it will project to zero. It's then easy to see that the tensor part must be equal to
\be{tenspart}
\frac{(YZ)X_M-(XZ)Y_M}{(XZ)^{1/2}(XY)^{1/2}(YZ)^{1/2}} \ ,
\ee
where the relative coefficient was fixed from the transversality constraint. We now have to project the tensor part into the physical space, i.e.~multiply by $\partial Z^M/\partial z^\mu$. We find that
\be{prZXY}
X_M\rightarrow (x-z)_\mu \ \ \ \text{and} \ \ \ \ Y_M\rightarrow(y-z)_\mu \ ,
\ee
therefore the expression projects into
\be{tensprojZ}
\frac{(x-z)_\mu\vert y-z\vert^2-(y-z)_\mu\vert x-z\vert^2}{\vert x-z\vert\vert y-z\vert\vert x-y\vert} \ , 
\ee
or in a nicer form 
\be{nicer}
\frac{\vert y-z\vert\vert x-z\vert}{\vert x-y\vert}\left(\frac{(x-z)_\mu}{\vert x-z\vert^2} -\frac{(y-z)_\mu}{\vert y-z\vert^2} \right)\equiv R_\mu(x,y\vert z) \ .
\ee
The quantity $R_\mu$ transforms correctly under CT. It turns out that this is the only indexed object for three points with this property ($I_{\mu\nu}$ is not useful here since it has two indices at different points). For spin $l$ fields the above formula is generalized to
\be{genmoreind}
\langle \phi_1(x)\phi_2(y)\phi_{3\mu\nu\lambda\ldots}(z)\rangle\propto\text{scalar part}\times(R_\mu R_\nu R_\lambda\cdots -\text{traces}) \ .
\ee
We see that the 3pt function is again completely fixed up to an arbitrary constant. 

An important special case is 3pt functions of two scalars with the current $J_\mu$ and the stress tensor $T_{\mu\nu}$:
\be{curstress3pf}
\langle \phi_1(x)\phi_2(y)J_{\mu}(z)\rangle\ , \ \ \ \langle \phi_1(x)\phi_2(y)T_{\mu\nu}(z)\rangle\ .
\ee
The above construction gives conformally invariant candidate expressions for these 3pt functions. What if we impose the conservation condition? For 2pt functions conservation was automatic, but here it is not so. In fact, the candidate 3pt functions are conserved if and only if the scalars have equal dimensions, $\Delta_1=\Delta_2$. Intuitively this happens because the 3pt function must satisfy the Ward identities, which relate it to the 2pt function. As we have seen, the 2pt function is non-zero if and only if $\Delta_1=\Delta_2$. The conclusion is that the coupling of the stress tensor and conserved currents to two scalar primaries of unequal dimensions must vanish.

In this lecture we learned to dominate the kinematics of the conformal group. Even though it gets us a long way, by itself it is not enough to solve a theory. In the next lecture we will introduce dynamics in the guise of Operator Product Expansion (OPE).

\section{Appendix: an elementary property of CTs}
It's good to know the following simple property of conformal transformations: they map circles to circles (considering straight lines as circles with infinite radius).

On the lightcone this is completely obvious. The Euclidean circle is represented on the lightcone by a set of points $X$ satisfying $X.X_0=c$ where $X_0$ is a fixed vector and $c$ a constant. A Lorentz transformation maps this to $X.X_0'=c$, which is another circle.


This property of CT is well known and useful in classical geometry. Here is an amusing geometrical problem which becomes easy if one uses it but would be tricky to solve otherwise: Given a circle $\gamma$ and two points $A, B$ outside of it, construct a circle tangent to $\gamma$ that passes through $A \ \text{and} \ B$. 

\emph{Idea of solution:} Apply an inversion with respect to any point on $\gamma$. The circle is now mapped to a straight line. In these coordinates, the problem is reduced to solving a quadratic equation. Then map back.

\section*{Literature}

The projective null cone idea is due to Dirac, and was used by Mack and Salam, Ferrara et al, Siegel and others. More recently it was used by Cornalba, Costa and Penedones \cite{Cornalba:2009ax} in its essentially modern form. It was then discussed by Weinberg \cite{Weinberg:2010fx} (who apparently rediscovered it independently of all the previous literature). This formalism is also developed further in our work \cite{Costa:2011mg}.
  
For the pedestrian approach to conformal correlators (as opposed to the null cone) see e.g.~Osborn and Petkou \cite{Osborn:1993cr}.

The classic 1970 paper by A.M.~Polyakov is \cite{Polyakov:1970xd}. It gave birth to CFTs, together with Mack and Salam's \cite{Mack:1969rr}.


\chapter{Radial quantization and OPE}
\label{lecture3}
\section{Radial quantization}

Until now we talked about correlation functions of local operators in the statistical mechanics sense, thinking of them as some sort of averages. In this lecture we will develop a parallel point of view, based on the Hilbert space and on quantum mechanical evolution. This will be quite useful for understanding all the questions related to the Operator Product Expansion (OPE). 

\subsection{General remarks on quantization}
Generally, the Hilbert space construction in QFT is linked to the choice of spacetime foliation:
\begin{figure}[h]
\begin{center}
\includegraphics[width=3cm]{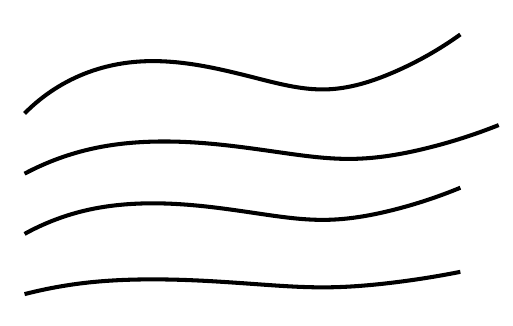}
\caption{$D$-dimensional spacetime foliated by $(D-1)$-dimensional surfaces (leafs).}
\end{center}
\end{figure}

 Each leaf of the foliation becomes endowed with its own Hilbert space. We create $in$ states $\vert\Psi_{in}\rangle$ by inserting operators or throwing particles in the past of a given surface. Analogously we deal with $out$ states when we insert operators or measure particles in the future. The overlap of an $in$ and $out$ state living on the same surface 
\be{cor1}
\langle \Psi_{out}\vert\Psi_{in}\rangle \ .
\ee
is equal to the correlation function of operators which create these $in$ and $out$ states (or to the $S$-matrix element if we deal with particle scattering).
\begin{figure}[h]
\begin{center}
\includegraphics[width=4cm]{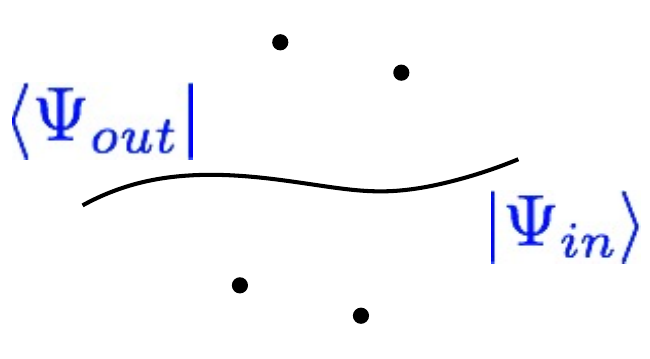}
\caption{Operator insertions in the past and present create Hilbert space states, whose overlap is equal to the correlation function involving these operators.}
\end{center}
\end{figure}

 If the in and out states live on different surfaces, and nothing is inserted in between, there will be a unitary evolution operator $U$ connecting the two Hilbert states and the same correlation function will be equal to:
\be{unU}
\langle \Psi_{out}\vert U\vert\Psi_{in}\rangle \ .
\ee
\begin{figure}[h]
\begin{center}
\includegraphics[width=4cm]{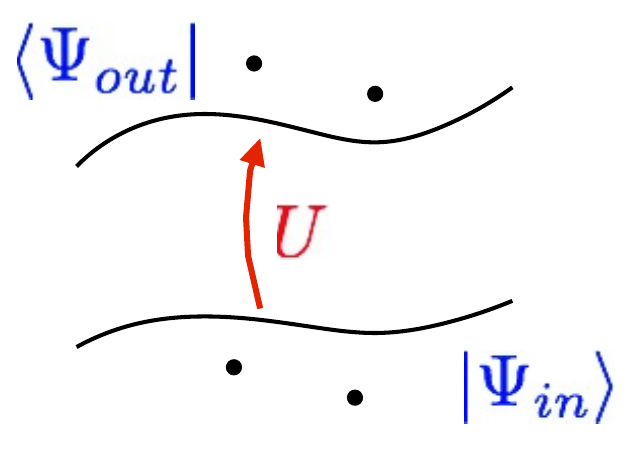}
\caption{If we choose to consider states living on different surfaces, then we must insert the evolution operator to evaluate the correlation function.}
\end{center}
\end{figure}

The above remarks are very general. In practice one likes to choose foliations that respect the symmetries of the theory. For example, in Poincar\'e invariant theories, we usually choose to foliate the space by surfaces of equal time. This is convenient, since the Hamiltonian $P^0$ moves us from one such surface to the other. The evolution operator $U$ will be simply given by exponentiating the generator $P^0$
\be{expon}
U=e^{i P^0\Delta t} \ .
\ee
Also, since all surfaces are related by a symmetry transformation, the Hilbert space is the same on each surface. The states living on these surfaces can be characterized by their energies and momenta
\be{enmom}
P^\mu\vert k\rangle=k^\mu\vert k\rangle \ .
\ee

\begin{figure}[h!]
\begin{center}
\includegraphics[width=3.5cm]{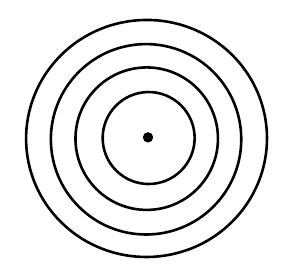}
\caption{Foliation by spheres $S^{D-1}$ all centered at the origin $x=0$ of the Euclidean $D$-dimensional space.}
\label{radfol}
\end{center}
\end{figure}

Another choice of foliation turns out more convenient in CFT, where we want to describe states created by insertions of local operators. Namely, we will use foliations by $S^{D-1}$ spheres of various radii, as in figure \ref{radfol}.
This is called radial quantization. We will assume that the center of the spheres is located at $x=0$, but of course quantizing with respect to any other point should give the same correlators. Similar arbitrariness is also present in the usual, ``equal time", quantization, since we have to fix a timelike time vector and different choices give rise to different but equivalent quantizations.

In radial quantization, the generator that moves us from one surface to another is the dilatations generator $D$, and it will play the role of the Hamiltonian:
\be{dilatU}
U=e^{i D\Delta \tau} \ ,
\ee
where $\tau=\log r$. The states living on the spheres will be classified according to their scaling dimension
\be{scaledimstat}
D\vert\Delta\rangle=i\Delta\vert\Delta\rangle \ , 
\ee
and their $SO(D)$ spin $l$
\be{spinstat}
M_{\mu\nu}\vert \Delta,l\rangle_{\{ s\} }=\left(\Sigma_{\mu\nu}\right)^{ \ \{t\}}_{\{s\}}\vert \Delta,l\rangle_{\{ t \} } \ ,
\ee
since the only generators that commute with $D$ are the angular momentum operators $M_{\mu\nu}$. The finite dimensional matrices $\Sigma_{\mu\nu}$ acting on the spin indices are nontrivial only for fields of non-zero spin $l$.

\subsection{Algebra action on quantum operators}

The next step is to understand how the operators transform. In the previous lectures we discussed how operators transform under finite Conformal Transformations (CT), from the statistical mechanics point of view. For example for scalars 
\be{ctscalar}
x\rightarrow x' \ , \ \ \  \phi(x)\rightarrow\phi(x')=\frac{1}{b(x)^\Delta}\phi(x) \ .
\ee
The above relation in terms of the correlation functions is understood as
\be{corctscalar}
\langle \phi(x')\ldots\phi(y')\rangle=\frac{1}{b(x)^\Delta}\cdots\frac{1}{b(y)^\Delta}\langle \phi(x)\ldots\phi(y)\rangle \ , 
\ee
or equivalently
\be{corctscalar2}
\langle b(x)^\Delta\phi(x')\ldots b(y)^\Delta\phi(y')\rangle=\langle\phi(x)\ldots\phi(y)\rangle \ .
\ee
Let us consider now the infinitesimal form of the transformation
\be{infform}
x'^\mu=x^\mu+\epsilon^\mu(x) \ \ \ \text{and} \ \ \ b(x)=1+\partial_\mu\epsilon^\mu \ .
\ee
Here for each group generator $G$ there will be its own $\epsilon_\mu$. Expanding the fields to first order in $\epsilon$, we can write
\be{fieldexpe}
b(x)^\Delta\phi(x')=\phi(x) + \epsilon(G\phi(x)) \ ,
\ee
where $G$ should be understood as the action of a generator on the field
\be{actionGfield}
G\phi(x)=(\Delta\partial_\mu\epsilon^\mu)\phi(x)+\epsilon^\mu\partial_\mu\phi(x) \ .
\ee
 Using the above expressions, we can write the correlation functions conformal invariance property in infinitesimal form
\be{corfunctionsinfinitesimal}
\langle G\phi(x_1)\phi(x_2)\phi(x_3)\ldots\rangle+\langle\phi(x_1)G \phi(x_2)\phi(x_3)\ldots\rangle+\ldots=0 \ .
\ee
In QFT, the action of the generators on quantum fields will be given by the same formula but we will write the action $G\phi$ as the commutator $[G,\phi(x)]$. The vacuum averages of quantum mechanical operators are the correlation functions
\be{qmcorrfun}
\langle \phi(x_1)\ldots\phi(x_n)\rangle=\langle 0\vert\phi(x_1)\ldots\phi(x_n)\vert 0 \rangle \ ,
\ee
and for consistency the vacuum is supposed to be conformally invariant:
\be{vacCI}
G\vert 0\rangle=0 \ .
\ee

Since we know $\epsilon$'s for all generators, we can work out how the generators act on the fields:
\be{actgenall}
\begin{aligned}
&[P_\mu,\mathcal O(x)]=-i\partial_\mu\mathcal O(x) \ ,\\
&[D,\mathcal O(x)]=-i(\Delta+x^\mu\partial_\mu)\mathcal O(x) \ ,\\
&[M_{\mu\nu},\mathcal O(x)]=-i(\Sigma_{\mu\nu}+x_\mu\partial_\nu-x_\nu\partial_\mu)\mathcal O(x) \ ,\\
&[K_\mu,\mathcal O(x)]=-i(2x_\mu \Delta+2x^\lambda\Sigma_{\lambda\mu}+2x_\mu(x^\rho\partial_\rho)-x^2\partial_\mu)\mathcal O(x) \ ,
\end{aligned}
\ee
where $\Delta$ the scaling dimension of $\mathcal O(x)$ and $\Sigma_{\mu\nu}$ the finite dimensional spin matrices we encountered previously. Each commutator has a part coming from $\epsilon$, and an extra part that comes either from the expansion of $b(x)$, or the $\Sigma$ matrices.

Notice that specializing to the point $x=0$, we get
\be{specx=0}
[K_\mu,\mathcal O(0)]=0 \ .
\ee
This property is in fact the defining property of primary operators. From this property, together with the natural ones:
\beq
\begin{aligned}
&[P_\mu,\mathcal O(x)]=-i\partial_\mu\mathcal O(x) \ ,\\
&[D,\mathcal O(0)]=-i\Delta\mathcal O(0) \ ,\\
&[M_{\mu\nu},\mathcal O(0)]=-i\Sigma_{\mu\nu}\mathcal O(0)\ ,
\end{aligned}
\eeq
and the conformal algebra, Eqs. \reef{actgenall} follow.

\subsection{Examples of states in radial quantization. State-operator correspondence.}

Let us now go back to the radial quantization. We will generate states living on the sphere by inserting operators inside the sphere. Some simple cases are:

1. The vacuum state $\vert 0\rangle$ corresponds to inserting nothing. The dilatation eigenvalue, the ``radial quantization energy",  is zero for this state.

2. If we insert an operator $\mathcal O_{\Delta}$ at the origin $x=0$, the generated state $\vert\Delta\rangle=\mathcal O_{\Delta}(0)\vert0\rangle$ will have energy equal to the scaling dimension $\Delta$: 
\be{oporigin}
\begin{aligned}
D\vert\Delta\rangle&=D\mathcal O_{\Delta}(0)\vert 0\rangle=[D,\mathcal O_{\Delta}(0)]\vert 0\rangle+\mathcal O_{\Delta}(0)D\vert 0\rangle\\
&=i\Delta\mathcal O_{\Delta}(0)\vert 0\rangle\ =i\Delta|\Delta\rangle.
\end{aligned}
\ee

3. If we insert an operator $\mathcal O_{\Delta}(x)$ with $x\neq 0$, the resulting state  $\vert \Psi\rangle=\mathcal O_{\Delta}(x)\vert0\rangle$ is clearly not an eigenstate of the dilatation operator $D$, as dilatations will move the insertion point. This can also be seen from Eq. \eqref{actgenall}. This state is however a superposition of states with different energies. This statement becomes clear if we write
\be{evolexp}
\vert \Psi\rangle=\mathcal O_{\Delta}(x)\vert0\rangle=e^{iPx}\mathcal O_{\Delta}(0)e^{-iPx}\vert0\rangle=e^{iPx}\vert\Delta\rangle=\sum_n \frac{1}{n!}(iPx)^n\vert\Delta\rangle \ .
\ee
When the momentum operator $P_\mu$ acts on $\vert\Delta\rangle$, it raises the energy by one unit. Algebraically, this is a consequence of 
\be{praise}
[D,P_\mu]=iP_\mu \ .
\ee
Schematically
\be{schemP}
\vert\Delta\rangle\xrightarrow{P_\mu}\vert\Delta+1\rangle\xrightarrow{P_\nu}\vert\Delta+2\rangle\cdots \ .
\ee
These states, associated with the derivatives of the primary operator $\calO$, are called descendant states.

On the other hand, the operator $K_\mu$ lowers the dimension by 1, since
\be{Klower}
[D,K_\mu]=-iK_\mu \ ,
\ee
thus
\be{schmeK}
0\xleftarrow{K_\mu}\vert\Delta\rangle\xleftarrow{K_\nu}\vert\Delta+1\rangle\cdots \ .
\ee
This allows us to justify the existence of primary operators, considered up to now as an axiom. Start with any local operator and keep hitting it with $K_\mu$.  Assuming that dimensions are bounded from below (as they are in unitary theories, see below), eventually we must hit zero, and this will give us a primary.

We saw that the states generated by inserting a primary operator at the origin have a definite scaling dimension $\Delta$ and are annihilated by $K_\mu$. We can go backwards as well: given a state of energy $\Delta$ which is annihilated by $K_\mu$, we can construct a local primary operator of dimension $\Delta$. This is called state-operator correspondence: states are in one-to-one correspondence with local operators.

The proof is easy. To construct an operator we must define its correlation functions with other operators. Define them by the equation
\be{defcorO}
\langle\phi(x_1)\phi(x_2)\ldots\mathcal O_\Delta(0)\rangle=\langle0\vert \phi(x_1)\phi(x_2)\ldots\vert\Delta\rangle \ .
\ee
This definition can be shown to satisfy all the usual transformation properties dictated by CI.

\subsection{Cylinder interpretation}

Introduce radial coordinates $r>0$, $\vec n\in S^{D-1}$ on $\mathbb R^D$, as well as
\be{polar}
\tau=\log r,
\ee
which shifts under dilatations $r\rightarrow e^\lambda r$:
\be{tautrans}
\tau\rightarrow\tau+\lambda 
\ee
In terms of these coordinates, local operator correlation functions must take the form
\be{taucorrfunctions}
\langle\phi(r_1,\vec n_1)\phi(r_2,\vec n_2)\ldots\rangle=\frac{1}{r_1^{\Delta_1}}\frac{1}{r_2^{\Delta_2}}\cdots f(\tau_i-\tau_j,\{\vec n_i\}) \ ,
\ee
where the function $f$ can depend only on the differences $\tau_i-\tau_j$ and all the unit vectors $\vec n_i$. Indeed, the factors $1/r_i^{\Delta_i}$ already account for the scaling. What remains must be scale-invariant, so it can only depend on ratios of $r_i$'s, or differences of $\tau_i$'s.

This suggests the following picture: start from flat space and go to the ``cylinder" $S^{D-1}\times\mathbb R$ parametrized by $\tau$ and $\vec n$. We can define fields on the cylinder by
\be{cylfield}
\phi_\text{cyl}(\tau,\vec n)=r^\Delta \phi_\text{flat}(r,\vec n) \ ,
\ee
where $\phi_\text{flat}(r,\vec n)$ are the fields that live on the flat space. With this identification, the function $f$ is the correlation function of the fields on the cylinder 
\be{fcylid}
\langle\phi_\text{cyl}(\tau_1,\vec n_1)\phi_\text{cyl}(\tau_2,\vec n_2)\ldots\rangle=f(\tau_i-\tau_j,\{\vec n_i\}) \ .
\ee
This suggest that the dynamics on the cylinder is invariant under translations of $\tau$. This is not accidental of course. This construction is more than just rewriting. Actually, the metric of the cylinder is equivalent to the flat space metric by a Weyl transformation:
\be{weyleq}
ds_\text{flat}^2=dr^2+r^2d\vec n^2\ , \ \ \ \text{and} \ \ \ ds_\text{cyl}^2=d\tau^2+d\vec n^2=\frac{1}{r^2}ds_\text{flat}^2 \ . 
\ee
\begin{figure}[htbp]
\begin{center}
\includegraphics[width=7cm]{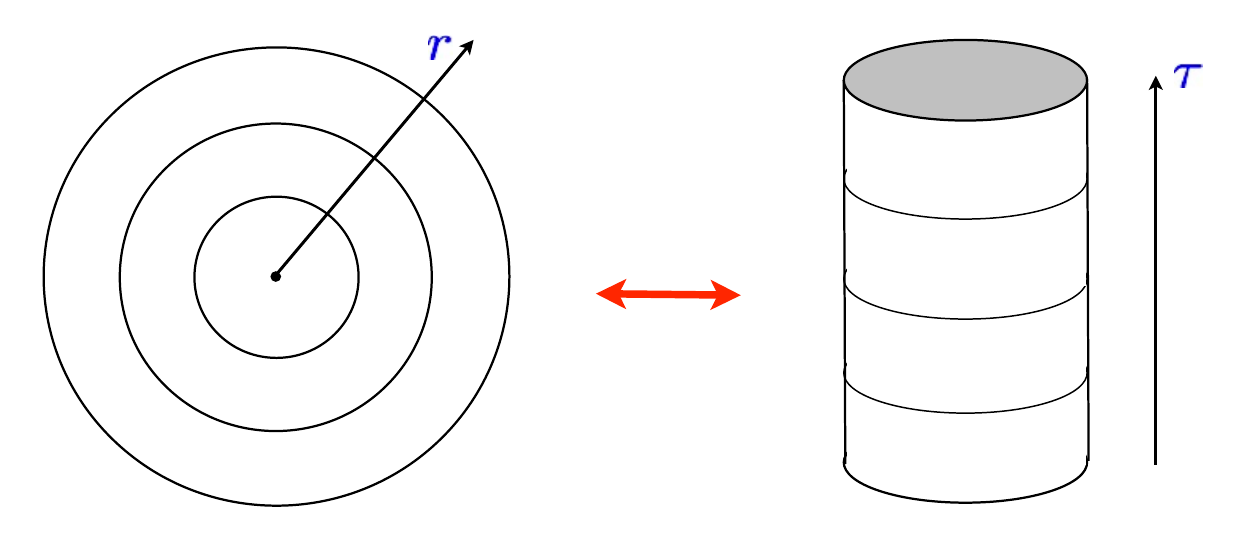}
\caption{Weyl transformation from ${\mathbb R}^D$ to the cylinder $S^{D-1}\times \bR$.}

\end{center}
\end{figure}

If the theory is conformal, then for any metric equivalent to the flat one by a Weyl transformation, the correlation functions are related by simple rescalling factors
\be{rescfact}
\langle \phi(x)\phi(y)\ldots\rangle_{g_{\mu\nu}=\Omega(x)^2\delta_{\mu\nu}}=\frac{1}{\Omega(x)^\Delta}\cdots\frac{1}{\Omega(y)^\Delta}\langle \phi(x)\phi(y)\ldots\rangle_\text{flat} \ .
\ee
This result might be a bit surprising, since we are considering a finite Weyl transformation. One might wonder if Weyl anomaly influences the correlation functions, but one can show that its contribution cancels out between the numerator and denominator (i.e. when one normalizes the correlator dividing by the partition function).

So, the cylinder field $\phi_\text{cyl}$ is not an artificial construct, but it is the very same field $\phi$ as in flat space, just its correlators are measured in a different geometry. As a concrete illustration of this non-trivial fact, suppose we do Monte-Carlo simulations of the 2d Ising model. We first put the theory on a flat, infinite lattice and adjust the temperature to its critical value. We calculate the correlation function of an operator inserted at two points. Then, we put the model on a cylindrical lattice, using the \emph{same} lattice action, and the \emph{same} temperature, and compute the correlation function of the same operator. These two correlators will be connected in the way we described above. 

\subsection{N-S quantization}

So far we have introduced two quantization pictures. The first one is the radial quantization. In this picture the in and out vacuum states sit at $0$ and $\infty$, so that the relation between them is not entirely obvious. The second one is the cylinder. In this picture the in and out states are at the two symmetric ends of the cylinder, and they can be related by the cylinder ``time" reflection $\tau\to-\tau$. To use the cylinder picture, one has to accept the equivalence relation \reef{rescfact}. 

We will now present the N-S (North-South pole) quantization picture which is in between the other two. It is almost as efficient as the cylinder, and in addition it takes place in flat space. Consider the following conformal group generator:
\be{cggen0}
K_1+P_1=i(2x_1(x^\mu\partial_\mu)-(1+x^2)\partial_1) \ .
\ee
It has two fixed points at $x_1=\pm 1$, $x_{2,\ldots,D}=0$ which we call N and S. The vector field takes us from one to the other. In this situation we can foliate our space using this generator as a Hamiltonian. 

\begin{figure}[htbp]
\begin{center}
\includegraphics[width=5cm]{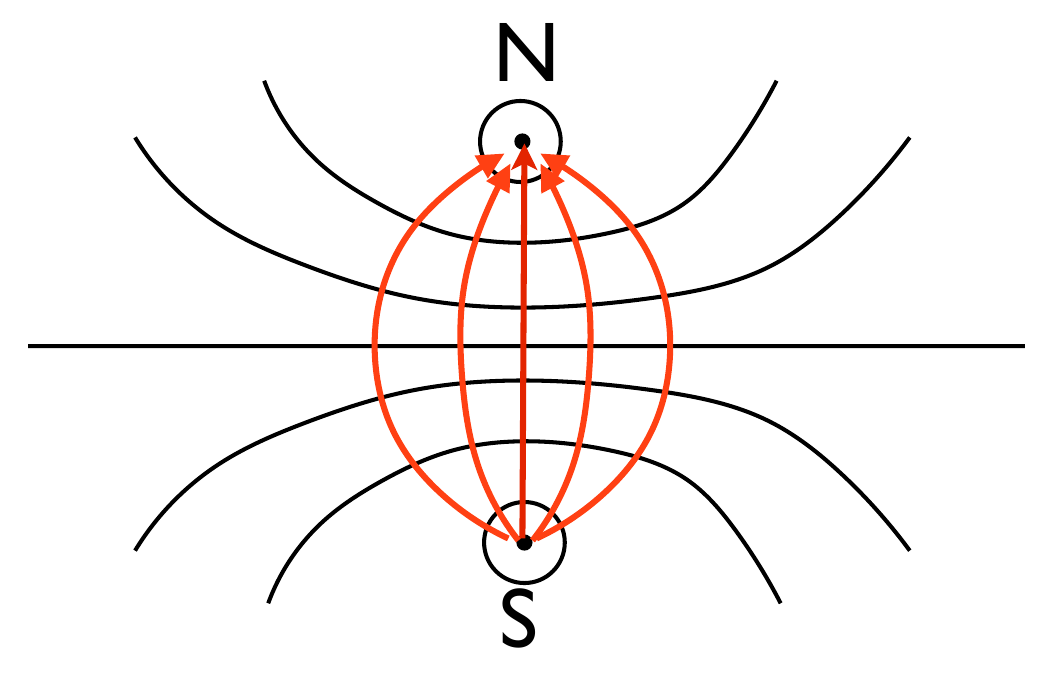}
\caption{NS foliation orthogonal to the $K_1+P_1$ vector field flow.}

\end{center}
\end{figure}

Such a quantization is completely equivalent to the radial quantization, because $K_1+P_1$ is conjugate to $D$ by $SO(D+1,1)$.\footnote{In the notation of the previous lecture, $K_1+P_1\sim J_{1,D+2}$, while $D\sim J_{D+1,D+2}$. These two generator are related by a rotation by $\pi/2$ in the $(D+1,1)$ plane.} In fact, we can map the above foliation to the radial one. To do that, apply an SCT to map the North pole to $\infty$:
\be{sctNpole}
x^\mu\to \frac{x^\mu-\alpha^\mu x^2}{1-2(\alpha x)+\alpha^2x^2} \ , \ x^\mu=\alpha^\mu\to\infty \ ,
\ee
followed by a translation to move the image of the South pole to $0$. 

The N-S quantization procedure will be very convenient for defining conjugate states. 

Let us add some operators below the line $x^1=0$. This will generate 
a state $\vert\Psi\rangle$ on $x^1=0$. If we apply the reflection transformation 
$\Theta: x^1\rightarrow -x^1$, this state is mapped onto $\langle\Psi\vert$. If we compute $\langle\Psi\vert\Psi\rangle$ in a unitary theory, this quantity is the norm of the state and has to be positive:
\be{normpsi}
\langle\Psi\vert\Psi\rangle>0 \ .
\ee
More generally, any $2n$-point functions with operators inserted in a $\Theta$-invariant way will have to be positive in a unitary theory. This property is called reflection positivity.

Conjugation of operators is therefore straightforward in the N-S quantization:
\be{theta}
[\phi(x)]^\dagger=\Theta(\phi(x))=\phi(\Theta x)
\eeq
(this is for scalar fields, for tensor fields $\Theta$ also changes signs of the $1$-components). Relation \reef{theta} may look strange, as the conjugate field lives at a point different from the original one, but it is in fact completely standard for the QFT Wick-rotated to the Euclidean signature. Indeed,
\beq
\phi(x,t)=e^{-iHt}\phi(x,0)e^{iHt}\ \to\ \phi(x,t_E)=e^{-Ht_E}\phi(x,0)e^{Ht_E}\,,
\eeq
and from here
\beq
\phi(x,t_E)^\dagger=e^{Ht_E}\phi(x,0)^\dagger e^{-Ht_E}=\phi(x,-t_E)
\eeq
provided that $\phi(x,0)$ is Hermitean.

\subsection{Conjugation in radial quantization}

The N-S quantization is conceptually nice, but to do calculations it's better to go back to the radial quantization. We have to understand how to define the conjugate state in the radial quantization picture. Since the two schemes, radial and N-S, are connected, we simply have to find the image of the reflection operator $\Theta$ under the transformation establishing the equivalence. Since this transformation maps $S\to0$ and $N\to \infty$, and since $\Theta$ interchanges $N$ and $S$ it is easy to guess that $\Theta$ is mapped to the inversion:
\be{thetainv} 
\Theta\rightarrow \text{inversion} \ \mathcal R \ .
\ee
The above implies that in the radial quantization, the conjugate of $\vert\Psi\rangle=\phi(x)\vert0\rangle$ will be
\be{conjpsirad}
\langle\Psi\vert=\langle0\vert[\phi(x)]^\dagger \ , \  [\phi(x)]^\dagger= r^{-2\Delta_\phi}\phi(\mathcal R x)\equiv \mathcal R[\phi(x)] \ .
\ee
This definition has the property that the correlation functions are $\mathcal R$-reflection positive (in a unitary theory):
\be{cfrefpos}
\langle 0\vert [\phi(y)]^\dagger[\phi(x)]^\dagger\ldots\phi(x)\phi(y)\vert 0\rangle>0 \ .
\ee

From this rule we can establish conjugation properties of algebra generators. In fact the operators $K_\mu$ and $P_\mu$ are conjugate by inversion:
\be{kpconj}
K_\mu=\mathcal R P_\mu \mathcal R  \ ,
\ee
(remember we said that SCT can be obtained by inversion followed by translation followed by inversion?). So we have:
\be{implrule}
(P_\mu\vert\Psi\rangle)^\dagger=\langle\Psi\vert K_\mu \ .
\ee
i.e.
\be{resultconj}
P_\mu=K_\mu^\dagger \ .
\ee

\subsection{2pt function in radial quantization}

For a concrete application of this formalism, consider the 2pt function of a scalar field
\be{2pfscalar}
\langle \phi(x_2)\phi(x_1)\rangle=\frac{1}{\vert x_1-x_2\vert^{2\Delta}} \ .
\ee
Since the action of $\phi(x)$ on the vacuum produces states with energies $E_n=\Delta+n$, we expect the radial quantization to give rise to following expansion:
\be{2pfscalarradial}
\langle 0|\phi(r_2,\vec n_2)\phi(r_1,\vec n_1)|0\rangle=\frac{1}{r_1^\Delta}\frac{1}{r_2^\Delta} \sum_n c_n e^{-E_n(\tau_2-\tau_1)} \ ,
\ee
where $\tau_2-\tau_1$ is the cylinder time interval: $e^{-(\tau_2-\tau_1)}=r_1/r_2$. 

To see that this is indeed true, let us rewrite \reef{2pfscalar} as
\be{expcalc2pf}
\langle \phi(r_2,\vec n_2)\phi(r_1,\vec n_1)\rangle=\frac{1}{r_1^\Delta}\frac{1}{r_2^\Delta}\frac{(r_1/r_2)^\Delta}{\bigl| 1-2\frac{r_1}{r_2}(\vec n_1 .\vec n_2)+\bigl(\frac{r_1}{r_2}\bigr)^{2} \bigr|^\Delta} \ .
\ee
Expanding the fraction in power of $r_1/r_2$, we get
\be{expfracr1r2}
\frac 1{\bigl| 1-2\frac{r_1}{r_2}(\vec n_1 .\vec n_2)+\bigl(\frac{r_1}{r_2}\bigr)^{2} \bigr|^\Delta}=\sum_n c_n \left(\frac{r_1}{r_2}\right)^n \ ,
\ee
where $c_n=P_n(\vec n_1.\vec n_2)$ will be some polynomials of $\vec n_1.\vec n_2$. This form agrees with \reef{2pfscalarradial}.

In fact, radial quantization can also be used to compute the coefficients $c_n$ independently. To accomplish that, let us write 
\be{actphi10right}
\phi(x_1)\vert0\rangle=e^{iPx_1}\vert\Delta\rangle \ .
\ee
We now have to understand the action of $\phi(x_2)$ on $\langle0\vert$. We showed that $[\phi(x)]^\dagger=r^{-2\Delta_\phi}\phi(\mathcal R x)$, thus
\be{actphi20left}
\langle0\vert\phi(x_2)=r_2^{-2\Delta_\phi}\langle0\vert[\phi(\mathcal R x_2)]^\dagger=r_2^{-2\Delta_\phi}\langle\Delta\vert e^{-iK \mathcal R x_2}\ .
\ee
So the correlation function becomes:
\be{fin2pf}
\langle 0\vert\phi(x_2)\phi(x_1)\vert 0\rangle=r_2^{-2\Delta_\phi}\langle\Delta\vert e^{-iK \mathcal R x_2}e^{iPx_1}\vert\Delta\rangle \ .
\ee
We can now expand the exponentials and get
\be{expexp}
r_2^{-2\Delta_\phi} \sum_{N}\langle\Delta\vert \frac{(-i K\mathcal R x_2)^N}{N!}\frac{(iPx_1)^N}{N!}\vert\Delta\rangle\ ,
\ee
where the cross terms with unequal number of powers of $K$ and $P$ will give zero matrix elements (since, as usual, states of unequal energy are orthogonal). We finally get
\be{expexp2}
\langle 0\vert\phi(x_2)\phi(x_1)\vert 0\rangle=\frac{1}{r_1^\Delta r_2^\Delta}\sum_{N}\langle N,\vec n_2\vert N,\vec n_1\rangle\left( \frac{r_1}{r_2}\right)^{\Delta+n} \ ,
\ee
where we denoted 
\be{stateNn}
\vert N,\vec n\rangle=\frac{1}{N!}(P_\mu \vec n^\mu)^N\vert\Delta\rangle\ . 
\ee
We can see that radial quantization relates the coefficients $c_n$ to certain matrix elements. These matrix elements can be evaluated purely algebraically, using conformal algebra. E.g.~for $N=1$ we get:
\be{N=1} 
\begin{aligned}
\langle \Delta\vert K_\mu P_\nu\vert\Delta\rangle&=\langle \Delta\vert [K_\mu, P_\nu]\vert\Delta\rangle+\langle \Delta\vert  P_\nu K_\mu\vert\Delta\rangle \\
&=\langle \Delta\vert 2i(D\delta_{\mu\nu}-M_{\mu\nu})\vert\Delta\rangle\\
&=\Delta\delta_{\mu\nu}\langle \Delta\vert\Delta\rangle=\Delta\delta_{\mu\nu} \ ,
\end{aligned}
\ee
where we used $K_\mu\vert\Delta\rangle=0$, and
\be{since} 
M_{\mu\nu}\vert\Delta\rangle=0 \ ,
\ee
for scalar fields. 

For $N>1$, we have to calculate an expression with the following structure
\be{explike}
\langle \Delta \vert K K K \ldots P P P\vert\Delta\rangle \ .
\ee
The evalutation proceeds along the same lines as above. We simply commute the $K$ operators until they hit the state $\vert\Delta\rangle$ and annihilate it. 

Proceeding this way, one can show order by order that the matrix elements agree with what expanding the known 2pt function predicts. Of course that's not the way to compute the 2pt function in practice, as fixing it from conformal kinematics is much simpler. But this check does show that radial quantization works, and in more complex cases we will be able to use it to go beyond kinematics. The first example of this is in the next section.
 
\section{Unitarity bounds}
 
A famous result most easily obtained by radial quantization for 2pt functions is the 
\emph{unitarity bound}: the dimension of a symmetric traceless primary field in a unitary theory must be above a minimal allowed value, $\Delta\ge \Delta_{min}(l)$. Depending on the spin, the minimal value is as follows:
\be{deltamin}
\begin{aligned}
\Delta_{min}(l)&=l+D-2\ , \ \ \ \text{if} \ \ \ l=1,2,3,\ldots \ \ \ \text{and} \\ 
\Delta_{min}(0)&=\frac{D}{2}-1\ .
\end{aligned}
\ee
Similar bounds exist for other $SO(D)$ representations, like antisymmetric tensors or fermions.

For the proof, consider the following matrix 
\be{matelA}
A_{\nu\{t\},\mu\{s\}}=_{\{t\}}\!\!\langle\Delta,l\vert K_\nu P_\mu\vert\Delta,l\rangle_{\{s\}} \ ,
\ee
where $\vert \Delta,l\rangle$ is a state created by inserting an operator of dimension $\Delta$ and spin $l$, and $\{s\}$ are the spin indices. In a unitary theory, this matrix must have only positive eigenvalues. Indeed, for a negative eigenvalue $\lambda<0$ with $\xi_{\mu,\{s\}}$ the corresponding eigenvector, the state $\vert\Psi\rangle=\xi_{\mu,\{s\}}P_\mu\vert\{s\}\rangle$ would have a negative norm: 
\be{contradiction}
\langle \Psi\vert\Psi\rangle=\xi^\dagger A\xi=\lambda\xi^\dagger\xi<0 \ .
\ee

Now using
\be{comrecunbounds}
[K_\nu,P_\mu]\propto i(D\delta_{\mu\nu}-M_{\mu\nu}) \ 
\ee
we notice that the eigenvalues of $A$ will get two contributions. The first will be proportional to $\Delta$, whereas the second will  be eigenvalues of a Hermitian matrix that depends only on the spin:
\be{matB} 
B_{\nu\{t\},\mu\{s\}}=\langle \{ t\}\vert i M_{\mu\nu}\vert \{s\}\rangle \ ,
\ee
The condition that all $\lambda_A\ge0$ will be this equivalent to
\be{eigenB}
\Delta\ge\lambda_{max}(B)\ , \ \ \ \text{where} \ \ \ \lambda_{max}(B) \ \text{is the maximum eigenvalue of $B$} \ .
\ee
The computation of the eigenvalues of $B$ is most easily done by using the analogy with how spin-orbit interaction is treated in quantum mechanics. Let us write the action of the operator $M_{\mu\nu}$ on the space of $\{s\}$ and $\{t\}$ in the following way
\be{SOB}
-i M_{\mu\nu}=-\frac{1}{2}(V^{\alpha \beta})_{\mu\nu}(M_{\alpha \beta})_{\{s\},\{t\}} \ ,
\ee
where the generator $(V^{\alpha b})_{\mu\nu}$ in the vector representation is given by
\be{vecV}
(V^{\alpha\beta})_{\mu\nu}=i(\delta^\alpha_\mu\delta^b_\nu-\delta^\alpha_\nu\delta^b_\mu) \ .
\ee
Let us compare the above with the standard problem in quantum mechanics, where we have to calculate the eigenvalues of 
\be{QMSO}
L^i \cdot S^i \ .
\ee
The operator $S^i$ is the analogue of $M_{\alpha \beta}$, they both act in the space of spin indices.
On the other hand $L^i$ is the analogue of $V^{\alpha \beta}$. The coordinate space in which $L_i$ acts is replaced by a vector space in which $V^{\alpha \beta}$ acts.

In QM, the diagonalization of \eqref{QMSO} is easily performed using the identity 
\be{idQMSO}
L^i \cdot S^i=\frac{1}{2}[(L+S)^2-L^2-S^2] \ .
\ee
Indeed, the operators $S^2$ and $L^2$ are Casimirs so their eigenvalues $s(s+1)/2$ and $l(l+1)/2$ are known, while $(L+S)^2$ is the Casimir of the tensor product representation $l\otimes s$ and so its eigenvalues are $j(j+1)/2$, $j=|l-s|,\ldots l+s$. 

In our case we have a spin representation $l$, a vector representation $V_{l=1}$ and a tensor representation $R'$ that occurs for $R'\in V\otimes l$. Thus, the maximal eigenvalue of $B$ will be given by 
\be{epcas}
\lambda_{max}(B)=\frac{1}{2}[\text{Cas}(V_{l=1})+\text{Cas}(l)-\min\text{Cas}(R')]\,.
\ee
For $l\ge 1$ the relevant representation $R'$ is the spin $l-1$ one. Using the $SO(D)$ Casimir value $l(l+D-2)$, we get $\lambda_{max}=l+D-2$, giving the stated unitarity bounds.

One may wonder if the bound could be strengthened by considering the states involving more $K$'s and $P$'s. This indeed happens for $l=0$ at level 2, where imposing that 
\be{lev2}
A_{\mu'\nu',\mu\nu}=\langle \Delta\vert K_{\mu'} K_{\nu'} P_\nu P_\mu \vert\Delta\rangle .
\ee
is positive definite, one derives
\beq
\Delta\ge \frac{D}{2}-1\,. \
\eeq

It turns out however that higher levels are not needed for spin $l\ge 1$, and that levels higher than 2 are not needed for scalars, i.e. the constraints we saw above are necessary and sufficient to have unitarity at all levels. 

\section{Operator Product Expansion (OPE)}

The idea of OPE is hopefully familiar from the usual QFT: it says that we should be able to replace a product of two local operators, in the limit when they become very close to each other, by a series of operators inserted at the midpoint. This logic still hold in CFT. Moreover, in CFTs the OPE acquires additional and very powerful properties thanks to a connection with the radial quantization.

\begin{figure}[htbp]
\begin{center}
\includegraphics[width=5cm]{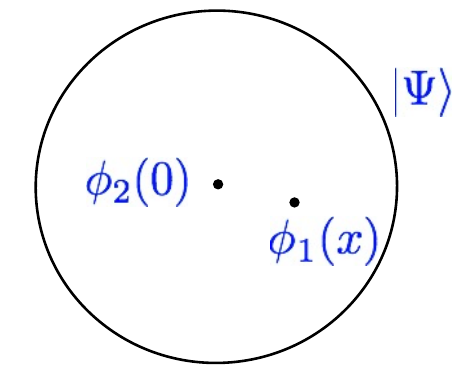}
\caption{Two operators inserted inside a sphere generate a state on the sphere.}
\label{radOPE}
\end{center}
\end{figure}

Let us first of all \emph{derive} OPE using radial quantization.
Suppose we have two operators inserted inside a sphere, as in figure \ref{radOPE}.

They generate a state $\phi_1(x)\phi_2(0)\vert 0\rangle=\vert\Psi\rangle$ on the surface of the sphere. This state will have an expansion into the basis of energy (i.e.~dilatation) eigenstates:
\be{exppsienergy}
\vert\Psi\rangle=\sum_n c_n \vert E_n\rangle \ , \ c_n= c_n(x) \ .
\ee
By the state-operator correspondence, the states $|E_n\rangle$ will be in one-to-one correspondence  with operators that are either primaries or derivatives (descendants) of primaries. We can thus write
\be{exprprimdes}
\phi_1(x)\phi_2(0)\vert 0\rangle=\sum_{\mathcal O \ \text{primaries}}C_\calO(x,\partial_y)\mathcal O(y)\vert_{y=0}\vert0\rangle \ ,
\ee
where $C_\calO(x,\partial_y)$ is understood as a power series in $\del_y$. We have just proved the existence of an OPE! 

In usual QFT, the OPE is usually used only in the asymptotic short-distance limit.
The fact that in CFT the OPE can be understood as an expansion in a Hilbert space which is complete has an extremely important consequence: the OPE in CFTs is not just asymptotic, but it gives a convergent series expansion at finite point separation. The radius of convergence will be discussed later.

So far we have only proved the existence of OPE. We can learn more about the coefficients by requiring that they must be consistent with the conformal algebra. Let us focus on a particular term of the series
\be{parttermseries}
\phi_1(x)\phi_2(0)\vert 0\rangle=\frac{const.}{\vert x\vert^{k}}[\mathcal O(0)+\ldots]\vert 0\rangle +\text{ contributions of other primaries},
\ee
where the power $k$ in the denominator will be fixed by scaling, and \ldots\ stand for terms with derivatives of $\mathcal O$. To fix $k$, let us act with the operator $D$ on the above expression. We start from the LHS 
\begin{align}
D\phi_1(x)\phi_2(0)\vert 0\rangle&=i(\Delta_1+x^\mu\partial_\mu)\phi_1(x)\phi_2(0)\vert 0\rangle+i\Delta_2\phi_1(x)\phi_2(0)\vert 0\rangle \ \nn\\
&\stackrel{\text{using \reef{parttermseries}}}{=}i(\Delta_1+\Delta_2-k)\frac{const.}{\vert x\vert^{k}}[\mathcal O(0)+\ldots]+\ldots
\label{lhsOPE}
\end{align}
On the other hand, acting with $D$ on the RHS of \reef{parttermseries}, we see
\be{rhsOPE}
D\frac{const.}{\vert x\vert^{k}}[\mathcal O(0)+\ldots]\vert 0\rangle=\frac{const.}{\vert x\vert^{k}}[i \Delta_O\mathcal O(0)+\ldots]\vert 0\rangle \ ,
\ee
where the dilatation operator acts only on $\mathcal O$, since the prefactor is simply a c-number. By comparing \reef{lhsOPE}
with \reef{rhsOPE} we conclude that $k=\Delta_1+\Delta_2-\Delta_O$.

Let us now consider the descendant terms:
\beq
\phi_1(x)\phi_2(0)\vert 0\rangle=\frac{const.}{\vert x\vert^{k}}[\mathcal O(0)+cx^\mu\del_\mu O(0)+\ldots]\vert 0\rangle +\text{ contributions of other primaries}.
\eeq
The coefficient $c$ can be found by acting on both sides of this equation with $K_\mu$. Once again we consider first the LHS
\begin{align}
\label{abKlhs}
K_\mu\phi_1(x)\phi_2(0)\vert 0\rangle&=(\sim x_\mu+\sim x^2\partial_\mu)\phi_1(x)\phi_2(0)\vert 0\rangle \ ,\\
&\sim\frac{const. x_\mu}{\vert x\vert^{k}}[\mathcal O(0)+\ldots]
\end{align}
where in the first line we used $K_\mu\phi_2(0)=0$, since $\phi_2(0)$ is a primary, and in the second line substituted the OPE. 
Acting on the RHS we get:
\be{abKrhs}
\begin{aligned}
K_\mu\frac{const.}{\vert x\vert^{k}}&[\mathcal O(0)+c x^\nu\partial_\nu \mathcal O(0)+\ldots]\vert 0\rangle\\
&=\frac{const.}{\vert x\vert^{k}}[K_\mu\mathcal O(0)+c K_\mu x^\nu\partial_\nu \mathcal O(0)+\ldots]\vert 0\rangle\\
&\sim\frac{const.}{\vert x\vert^{k}}[c \Delta x^\mu \mathcal O(0)+\ldots] \ ,
\end{aligned}
\ee
where we used the fact that $K_\mu$ lowers the dimension by 1, so $K_\mu \mathcal O(0)=0$, whereas $K_\mu P_\nu\mathcal O(0)\sim \delta_{\mu\nu}\Delta \mathcal O(0)$. Matching the coefficients, we see that we can determine $c$. Analogously, the coefficients of the higher descendants can also be determined recursively. 

Conformal invariance thus fixes completely the function $C_\calO(x,\partial_y)$, up to a numerical prefactor $\lambda_O$:
\be{fixedc}
\sum_O \lambda_O C_{\mathcal O}(x,\partial_y)\mathcal O(y)\vert_{y=0}\vert 0\rangle \ , \ \ \ \text{with} \ \ \ \lambda_O \ \text{a free parameter.}
\ee

The above argument was useful conceptually, but in practice the function $C_\calO(x,\partial_y)$ is easier to compute by a different method. This method is based on the fact that we can use OPE to reduce $n$-point functions to $(n-1)$-point functions. Let's apply this to a scalar 3pt function. Using the OPE inside the correlator, we can write:
\be{3pfOPE}
\langle \phi_1(x)\phi_2(0)\Phi(z)\rangle=\sum_O \lambda_O C_{\mathcal O}(x,\partial_y)\langle \mathcal O(y)\vert_{y=0} \Phi(z)\rangle \ ,
\ee
Since we know that the 2pt function is diagonal, the only surviving term from the infinite sum will be for $O=\Phi$: 
\be{3pfOPE1}
\langle \phi_1(x)\phi_2(0)\Phi(z)\rangle=\lambda_\Phi C_{\Phi}(x,\partial_y)\langle \Phi(y)\vert_{y=0} \Phi(z)\rangle \ ,
\ee
Moreover, we know both the 2pt and 3pt functions entering the above equation:
\begin{gather}
\label{3pfknown}
\langle \phi_1(x_1)\phi_2(x_2)\Phi(x_3)\rangle=\frac{\lambda_\Phi}{(x_{12})^{\ldots}(x_{13})^{\ldots}(x_{23})^{\ldots}}\\
\langle\Phi(y)\Phi(z)\rangle=\frac{1}{\vert y-z\vert^{2\Delta_\Phi}} \ ,
\end{gather}
Now, expanding both sides around $x=0$ and fixing the coefficients term by term, we can determine the function $C_{\Phi}(x,\partial_y)$. Since we normalized the 3pt function using the same coefficient $\lambda_\Phi$, the function $C_{\Phi}(x,\partial_y)$ will depend only on the dimensions of involved fields (and the spacetime dimension).

\textbf{Exercise:} Matching the 3pt function to the OPE, show that for $\Delta_1=\Delta_2=\delta$ and $\Delta_\Phi=\Delta$,
\be{exercise}
C_\Phi(x,\partial_y)=\frac{1}{\vert x\vert^{2\delta-\Delta}}\left[1+\frac{1}{2}x^\mu\partial_\mu+\alpha x^\mu x^\nu\partial_\mu\partial_\nu+\beta x^2\partial^2+\ldots\right] \ ,
\ee
where
\be{exerciseab}
\alpha=\frac{\Delta+2}{8(\Delta+1)} \ \ \ \text{and} \ \ \ \beta=-\frac{\Delta}{16\left(\Delta-\frac{D}{2}+1\right)(\Delta+1)} \ .
\ee

\section*{Literature}

Radial quantization is usually discussed in courses on 2d CFT or on string theory. See e.g.~Joe Polchinski's ``String theory", Vol.1, where also the argument connecting OPE with the radial quantization is given. The presentation here is based on \cite{Pappadopulo:2012jk}

For the infinitesimal action of generators on fields, see Mack and Salam's classic work \cite{Mack:1969rr}.
  
  For the N-S quantization, see the classic paper by L\"uscher and Mack \cite{Luscher:1974ez}.
  Caution: heavily mathematically written, as befits a Communications in Mathematical Physics paper.
    
  For the unitarity bounds, see the classic work by Minwalla \cite{Minwalla:1997ka} where the necessary level 1 condition is explained and the analogy with the spin-orbit coupling is pointed out. To prove that the necessary condition is sufficient is much harder. In 4d this is done in a classic paper by Mack \cite{Mack:1975je}.
  
Recently, Penedones, Trevisani and Yamazaki \cite{Penedones:2015aga} and Yamazaki \cite{Yamazaki} pointed out the relevance of the old mathematical work of Jantzen \cite{jantzen_kontravariante_1977} to this problem. Apart from other things, this can be used to establish rigorously the unitarity bounds in any $D$ and for any representation.


\chapter{Conformal Bootstrap}
\label{lecture4}
\section{Recap}

By now we have learned quite a bit about the structure of CFTs. We have seen that:

$\bullet$ Any CFT is characterized by the spectrum of local primary operators, by which we mean the set of pairs $\{\Delta,\mathcal R\}$, where $\Delta$ is operator's scaling dimension, and $\mathcal R$ the representation of the $SO(D)$ under which it transforms. We showed that all other operators are obtained by differentiating the primaries; they are called descendants. 
We also showed that there is a one-to-one correspondence between the operators $\calO_\Delta$ and the states of a radially quantized theory, obtained by inserting the operator at the origin: $|\Delta\rangle=\calO_\Delta(0)|0\rangle$.
Finally, we showed that (in unitary theories) there exist lower bounds for the operators' dimensions
\be{unbound}
\Delta\ge\Delta_{min}(\mathcal R) \ 
\ee

$\bullet$ The 2pt functions of primaries are fixed. For example in the case of scalars
\be{2pfscal}
\langle\phi(x)\phi(y)\rangle=\frac{N}{\vert x-y \vert^{2\Delta}} \ ,
\ee
and analogously for higher spins. The normalization constant is usually fixed to $N=1$, which in the radial quantization language corresponds to the normalization choice $\langle\Delta\vert\Delta\rangle=1$. In radial quantization the above correlator can be computed as the matrix element
\be{2pfscalrq}
\langle\Delta\vert e^{-iK y/y^2}e^{i P x}\vert\Delta\rangle\ .
\ee
This matrix element can be evaluated just by using conformal algebra, which also explains why the 2pt function is fixed.

$\bullet$  The 3pt functions of primary operators are fixed up to a constant. For scalars, we saw that they are given by
\be{3pfscal}
\langle\phi_1(x_1)\phi_2(x_2)\phi_3(x_3)\rangle=\frac{\lambda_{123}}{|x_{12}|^{2\alpha_{123}}   
|x_{13}|^{2\alpha_{132}}|x_{23}|^{2\alpha_{231}}} \ ,\quad \alpha_{ijk}=\frac{\Delta_i+\Delta_j-\Delta_k}{2} \,,
\ee
where $\lambda_{123}$ a free parameter which one cannot rescale away once 2pt function normalization has been fixed. 

Analogously the 3pt function of two scalar and one spin-$l$ operator is fixed up to a constant. Using the methods of lecture \ref{lecture2}, one can also compute the most general 3pt function of three spin-$l$ operators. The number of tensor structures consistent with conformal symmetry is more than one in general, but it's always finite. The constants multiplying these tensors are parameters undetermined by kinematics alone. 

$\bullet$ As we discussed, the constant $\lambda_{123}$ in Eq.~\eqref{3pfscal} is the same as the one appearing in the conformally invariant OPE:
\be{CIOPE} 
\phi_1(x)\phi_2(0)=\sum_{\mathcal O \ \text{primaries}}\lambda_{12\mathcal O}\,C_{\mathcal O}(x,\partial_y)\mathcal O(y)\vert_{y=0} \ .
\ee
In our discussion, we inserted the operator $\mathcal O(y)$ at $y=0$, but it could be inserted in any point between $x \ \text{and} \ 0$. Of course, the coefficient functions $C_{\mathcal O}(x,\partial_y)$ will then have to be changed appropriately. This freedom will turn out to be useful later. 

Alternatively, the constants $\lambda_{12\calO}$ can be thought of as the coefficients appearing when the state $\phi_1(x)\phi_2(0)\vert 0\rangle$ is expanded into a complete basis of states:
\be{CIOPE2}
\phi_1(x)\phi_2(0)\vert 0\rangle=\sum_{\mathcal O}\lambda_{12\mathcal O}\left[ \mathcal O(0)+\ \text{descendants} \right] \ 
\ee
in the radial quantization scheme. A non-trivial feature of CFT is that the function $C_{\mathcal O}(x,\partial_y)$ is completely fixed, i.e. there is one constant $\lambda_{12\mathcal O}$ per conformal family.\footnote{As mentioned above, there are finitely many constants for the general external and internal representations.} The constants $\lambda_{ijk}$ are called interchangeably \emph{3pt function coefficients}, \emph{OPE coefficients}, or \emph{structure constants of the operator algebra}.

\begin{figure}[htbp]
\begin{center}
\includegraphics[height=3cm]{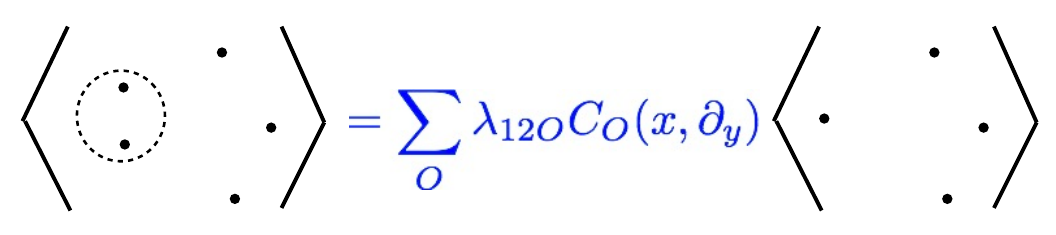}
\caption{Reducing a 5-point function to a sum of 4-point functions using the OPE.}
\label{5to4}
\end{center}
\end{figure}

$\bullet$ The set $\{\text{spectrum, OPE coefficients}\}$ is called \emph{CFT data}. With its knowledge, we can compute any $n$-point correlation function of the theory, recursively reducing it to $(n-1)$, $(n-2),$\ldots\ and finally to 2pt functions by using the OPE.\footnote{We can also stop at the 3pt functions since those are already fixed.} For example 5pt function are reduced to 4pt functions as in Fig.~\ref{5to4}.
Moreover, this expansion will be convergent. This becomes obvious if we view it as an expansion in the Hilbert space
\be{exphs} 
\vert \Psi_1\rangle=\sum_n c_n\vert n\rangle \ \ \  \rightarrow  \ \ \ \langle\Psi_2\vert\Psi_1\rangle=\sum_n c_n \langle \Psi_2\vert n\rangle \ .
\ee
Here $\vert \Psi_1\rangle$ and $\langle \Psi_2|$ are the states generated on the dashed sphere by operators inserted inside and outside. Then we expand $\vert \Psi_1\rangle$ into states generated by operators inserted at the center of the sphere (OPE), and get an expansion for the scalar product $\langle\Psi_2\vert\Psi_1\rangle$ which describes the correlation function. Expansions of scalar products are convergent for states of finite norm. In our case, the norms  $\langle\Psi_i\vert\Psi_i\rangle$ are finite since they correspond to some correlation function of operators inserted in an inversion-invariant way. The norms are finite as long as we can find a sphere separating strictly the internal two points from the rest. This gives the radius of convergence of the OPE.

\section{Consistency condition on CFT data}

As mentioned, given a CFT data, we can compute all the correlators in a theory. But does any random set of CFT data define a good theory? The answer is no. A consistency condition on the CFT data comes from studying the 4pt functions. Consider a scalar 4pt function in a generic point configuration:
\begin{center}
\includegraphics[height=2cm]{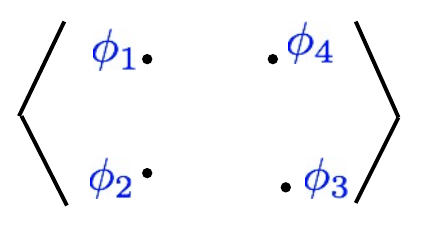}
\end{center}

To compute it via the OPE, we surround two of the operators, say $\phi_1$ and $\phi_2$ by a sphere and expand into radial quantization states on this sphere. Operationally this means that we are writing:
\be{schanOPE}
\begin{aligned}
\phi_1(x_1)\phi_2(x_2)&=\sum_{\mathcal O}\lambda_{12\mathcal O}C_{\mathcal O}(x_{12},\partial_y)\mathcal O(y)\Big\vert_{y=\frac{x_1+x_2}{2}}  \ ,\\ 
\phi_3(x_3)\phi_4(x_4)&=\sum_{\mathcal O}\lambda_{34\mathcal O}C_{\mathcal O}(x_{34},\partial_z)\mathcal O(z)\Big\vert_{z=\frac{x_3+x_4}{2}} \ .
\end{aligned}
\ee
Plugging the above into the 4pt function, we get
\be{4pfOPEschan}
\langle\phi_1(x_1)\phi_2(x_2)\phi_3(x_3)\phi_4(x_4)\rangle= \sum_{\mathcal O} \lambda_{12\mathcal O}\lambda_{34\mathcal O}
\bigl[C_{\mathcal O}(x_{12},\partial_y)C_{\mathcal O}(x_{34},\partial_z)\langle\mathcal O(y)\mathcal O(z) \rangle \bigr] . 
\ee
The functions in square brackets are completely fixed by conformal symmetry in terms of the dimensions of $\phi_i$ and of the dimension and spin of $\calO$ (since both the functions $C_{\mathcal O}$ and the correlator $\langle\mathcal O(y)\mathcal O(z) \rangle$ are fixed). These functions are called Conformal Partial Waves (CPW).
Diagramatically the expansion into conformal partial waves can be written as:
\newpage
~\begin{figure}[!h]
\centering
\includegraphics[scale=.8]{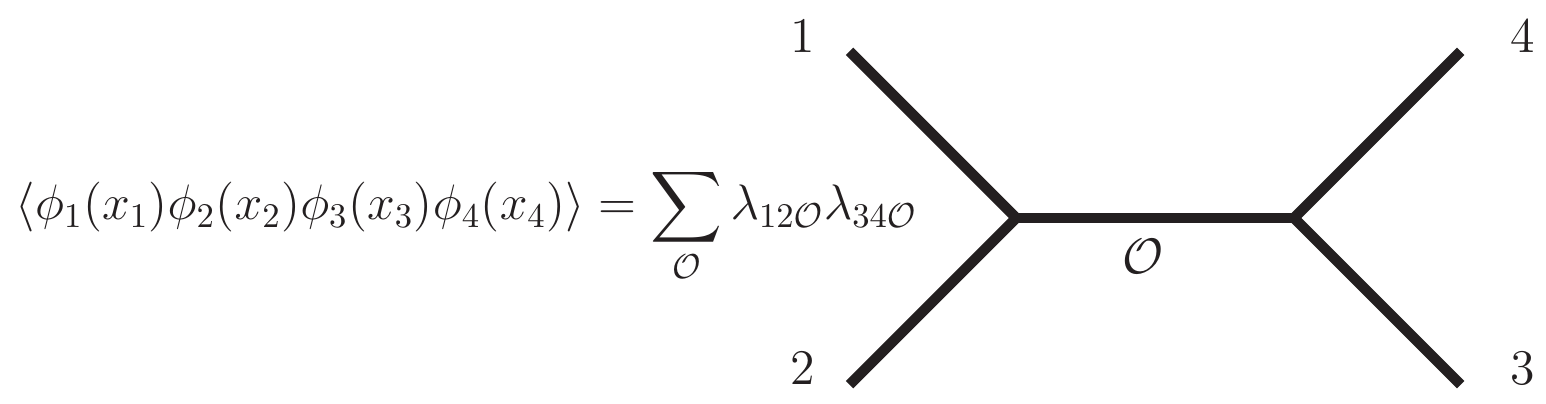}
\end{figure}

Notice that one should not confuse CPWs with perturbative Feynman diagrams. When expressing a QFT correlation function in terms of Feynman diagrams, we have to include three channels (s,t,u). Only one channel is sufficient to reproduce the 4pt function using the CPW expansion. Instead of summing the three channels, we have to impose that all of them agree. This is not automatic, and gives rise to a consistency condition.

Concretely, notice that we might have chosen to compute the same 4pt function by choosing a sphere enclosing the operators $\phi_1 \ \text{and} \ \phi_4$. This means that we would have chosen a different OPE channel, (14)(23) instead of (12)(34). We would have obtained a different CPW expansion, but the result should be the same:
~\begin{figure}[!h]
\centering
\includegraphics[scale=.8]{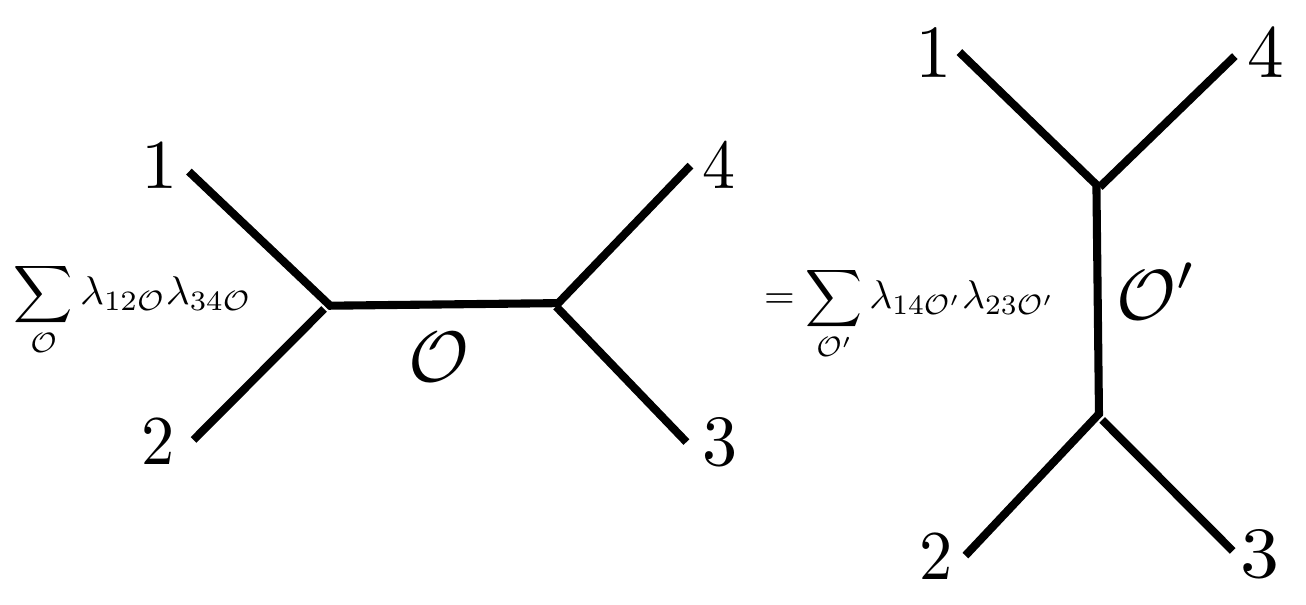}
\end{figure}

This condition is called in the literature interchangeably as ``OPE associativity'', ``crossing symmetry" and ``conformal bootstrap" condition. It has to be satisfied by any consistent CFT data. 

\section{Conformal Bootstrap}

\begin{figure}[!h]
\centering
\includegraphics[scale=.45]{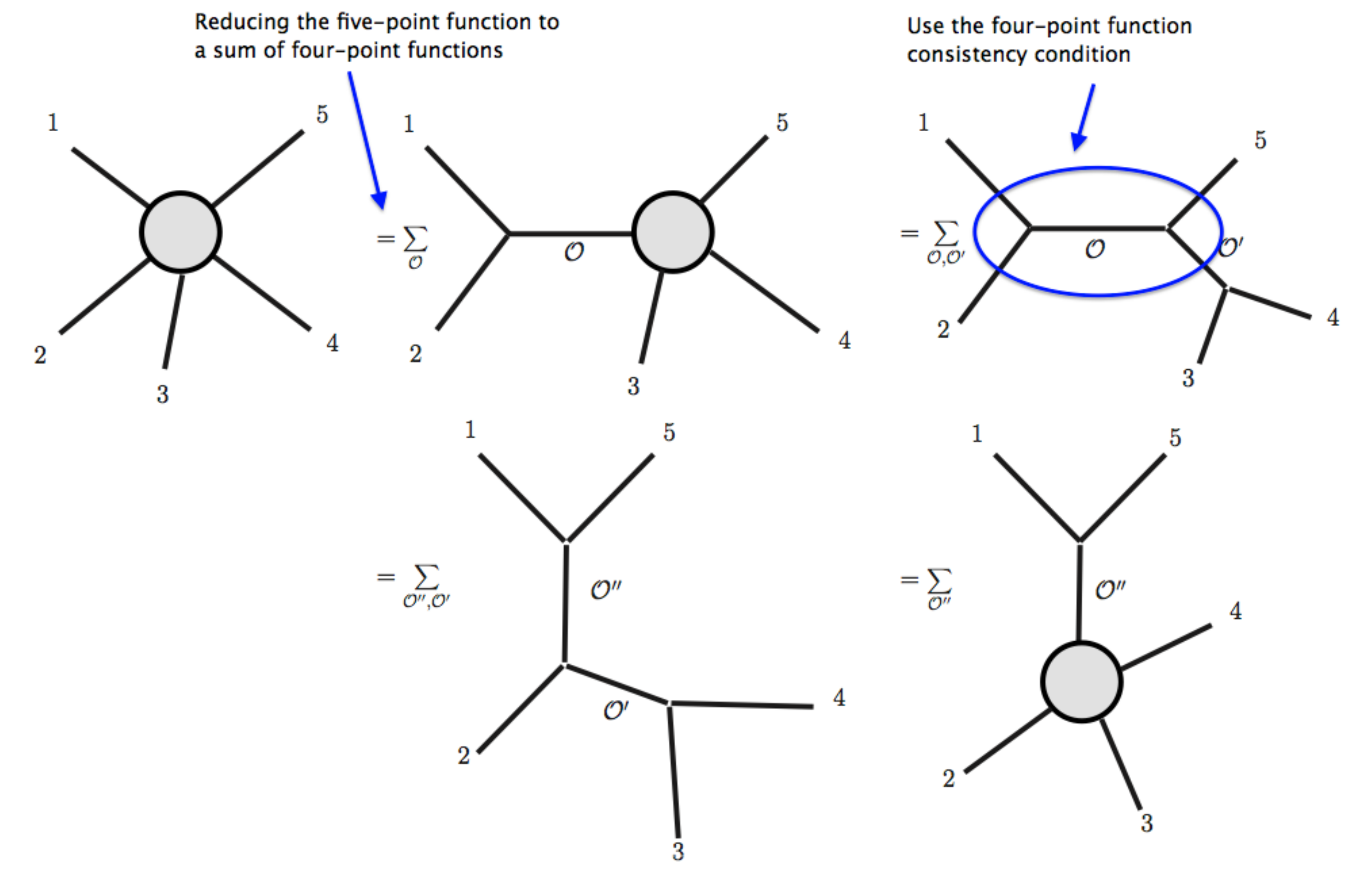}
\caption{For five-point functions, OPE in (12) and (15) channels should give equivalent expansions. This figure demonstrates schematically that this is not an additional constraint but follows from OPE associativity for 4pt functions.}
\label{fig:5pt}
\end{figure}

Once we impose OPE associativity on all 4pt functions, no new constraints will appear at higher $n$-point functions. For the 5pt function this is demonstrated in Figure \ref{fig:5pt}

This suggests that OPE associativity at the 4pt function level is close to a necessary and sufficient condition to have a consistent CFT. It's definitely necessary, although it may be not quite sufficient. Still, tentatively we may consider the following

{\bf Definition} [Ferrara, Grillo, Gatto 1973, Polyakov 1974, Mack 1977, \ldots]  {\it A CFT is a set of CFT data which satisfies the OPE associativity for all 4pt functions}.

This must not be taken too dogmatically. E.g.~we may have subsectors closed under the OPE, such as the singlet operators in a theory with a global symmetry. OPE associativity is not going to let us conclude that this is not the full theory. In $D=2$ modular invariance provides an extra constraint, but in $D\ge 3$ modular invariance has not been useful so far since there is no known way to express the partition function on $T^D$ in terms of CFT data.\footnote{It's easy on $S^{D-1}\times \bR$, but for $D>2$ this manifold does not allow modular invariance transformations.} If extra constraints are discovered in the future, one should not hesitate to update this definition.

Ideally, the above definition could be used to classify CFTs. This would be similar in spirit to how the finite-dimensional Lie algebras are classified starting from the Jacobi identity. More modestly, one can ask if it is possible to deduce from this definition some useful general CFT properties, beyond those already discussed.

This research direction is called ``conformal bootstrap''.  

\subsection{Success story in $D=2$}

In 2d CFT the power of the conformal bootstrap was demonstrated in the famous 1984 paper by Belavin, Polyakov and Zomolodchikov (BPZ). As discussed in lecture \ref{lecture1}, the 2d conformal algebra (Virasoro algebra) is infinite dimensional. The generators are called $L_n$ and $\bar L_n$. The finite dimensional subalgebra of global conformal transformations is embedded into it as follows:
\be{virasorgen}
\begin{aligned}
\underbrace{\ldots, L_{-3},L_{-2}}_{\text{extra raising operators}},&\underbrace{L_{-1},\quad L_0,\quad L_1}_{P_\mu \ \{D,M_{\mu\nu}\} \ \ K_\mu},\underbrace{L_2,L_3,\ldots}_{\text{extra lowering operators}}\\
\ldots\quad &\bar L_{-1},\quad \bar L_0,\quad \bar L_1 \ ,\quad\ldots
\end{aligned}
\ee
The generator $L_{-n}$ raises the state scaling dimension by $n$ units: $L_{-n}\vert \Delta\rangle=\vert\Delta+n\rangle$. A Virasoro primary field is annihilated by all lowering operators:
\be{virasorfieldprim}
L_{n}\vert\Delta\rangle=\bar L_{n}\vert\Delta\rangle=0 \quad(n\ge 1)\,.
\ee
This is stronger than the condition
\be{virasorcondK}
K_\mu\vert\Delta\rangle=0 \Leftrightarrow L_{1}\vert\Delta\rangle=\bar L_{1}\vert\Delta\rangle=0 \ ,
\ee
considered so far. The field satisfying the latter condition is called quasiprimary in $D=2$ CFT literature.
 
It would take us too long to explain the full 2d story. Very schematically, one first studies the unitarity conditions for the 2d infinite dimensional Virasoro algebra and finds that they are more restricted than in $D\ge 3$. In particular, the conditions depends on the value of a certain parameter $c$, called \emph{central charge}, which appears in the 2pt function of the canonically normalized stress tensor $T$:
\be{ccharge}
\langle T(x)T(y)\rangle\sim \frac{c}{(x-y)^4} \ .
\ee

1. For $c\ge 1$, one finds that the unitarity conditions are more or less the same as in the three-dimensional case. 

2. For $0<c<1$, the condition is much more restrictive. Namely, only a discrete sequence of values for $c$ is allowed
\be{cchargecond} 
c=1-\frac{6}{m(m+1)} \ , \ \text{with} \ m=3,4,5, \dots
\ee
Theories corresponding to such $c$ are called ``minimal models''. Moreover, one finds that for $c< 1$ only a finite discrete set of operator dimensions is allowed to appear:
\be{dimfixed}
\Delta_{r,s}=\frac{(r+m(r-s))^2-1}{2m(m-1)} \ , \ \text{where} \ 1\le s\le r\le m-1\ \ \text{integers} \ .
\ee
Notice that one primary multiplet in two dimensions splits into infinitely many quasiprimary multiplets. So one 2d primary is morally equivalent to infinitely many primaries in three dimensions. Thus we may have finitely many primaries in 2d (this happens in the minimal models), but we will always have infinitely many primaries in $D\ge 3$ dimensions. 

Once the dimensions are known and since there are finitely many primaries, the OPE associativity equations for the minimal models reduce to a problem of finite-dimensional linear algebra. Solving these equations one determines the OPE coefficients.

The simplest minimal model appears for $c=1/2$ and corresponds to the critical 2d Ising model. The Virasoro primary field content includes just the identity operator $\mathbb I$, the spin field $\sigma\ (\mathbb Z_2$-odd) and the energy field $\epsilon\ (\mathbb Z_2$-even), with dimensions
\be{is2dc1/2}
\Delta_\mathbb I=0 \ , \ \Delta_\sigma=\frac{1}{8} \ ,  \ \Delta_\epsilon=1 \ .
\ee
The OPEs, consistent with the $\mathbb Z_2$ symmetry of the model, are
\be{OPE2dising}
\begin{aligned}
&\sigma \times \sigma=\mathbb I+\lambda_{\sigma\sigma\epsilon}\epsilon\\
&\epsilon \times\epsilon=\mathbb I+\lambda_{\epsilon\epsilon\epsilon}\epsilon\\
&\sigma\times\epsilon=\lambda_{\sigma\sigma\epsilon} \sigma\ .
\end{aligned}
\ee
where $\lambda_{\sigma\sigma\epsilon}$ is determined by solving bootstrap equation, while $\lambda_{\epsilon\epsilon\epsilon}=0$ due to the Kramers-Wannier duality, a property of the critical Ising model specific to $D=2$.

While the minimal models have $c<1$, there are many CFTs with $c>1$ and infinitely many primaries. Conformal bootstrap equations are then difficult to solve even in $D=2$. One notable example where this has been done is the Liouville theory. 

\subsection{Tools for the bootstrap in $D\ge 3$ dimensions}

The key role will be played by the CPW decomposition, so let us rediscuss it more explicitly. Take for simplicity the 4pt function of scalars with the same scaling dimension $\Delta_i=d$ (not to be confused with the spacetime dimension). This is decomposed into CPW's as:
\begin{figure}[!h]
\centering
\includegraphics[scale=.8]{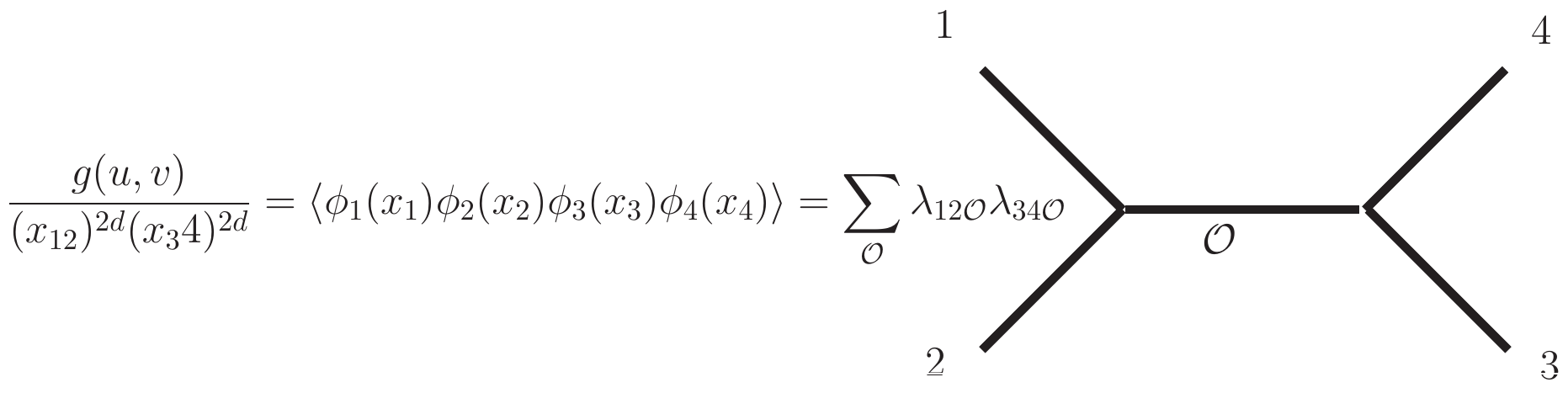}
\end{figure}

Actually, each CPW will have the same transformation properties under the conformal group as the 4pt function itself. This can be shown to follow from the conformal invariance of the OPE used to define CPW. In other words, we will have:
\be{4pfCPW2}
\langle\phi_1(x_1)\phi_2(x_2)\phi_3(x_3)\phi_4(x_4)\rangle=\sum_{\mathcal O} \lambda_{12\mathcal O}\lambda_{34\mathcal O}\frac{G_\mathcal O(u,v)}{(x_{12})^{2d}(x_{34})^{2d}} \ ,
\ee
where $G_\mathcal O(u,v)$ are called \emph{conformal blocks} (CBs).\footnote{In a part of literature the terms CPWs and CBs are used interchangeably, but here we distinguish them.} We have:
\beq
g(u,v)=\sum_{\mathcal O} \lambda_{12\mathcal O}\lambda_{34\mathcal O}{G_\mathcal O(u,v)}
\eeq

We need to gain some intuition about CBs. Since they depend only on $u$ and $v$, we can study them by using a conformal transformation to map the four points to some convenient positions. (Remember that conformal transformations leave $u$ and $v$ invariant).

Let us consider the following configuration: map the point $x_4\rightarrow\infty$. Then shift $x_1\rightarrow 0$, and do a rotation followed by a dilatation to put $x_3=(1,0,\ldots,0)$ (all these transformations leave $\infty$ invariant since there is only one $\infty$ point). So far we have fixed three points. Let us now do a rotation with respect to the $x_1$-$x_3$ axis, to put $x_2$ into the plane of this page. We use the $z$ complex coordinate in this plane:
\begin{center}
\includegraphics[height=3cm]{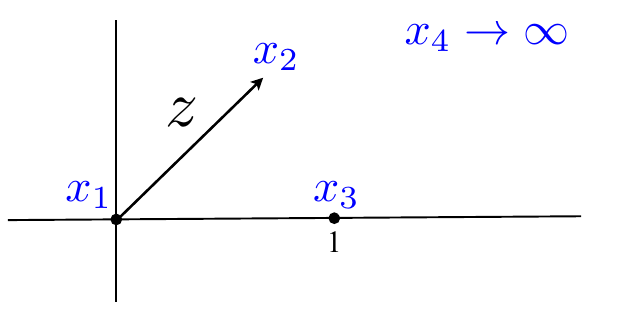}
\end{center}

The cross-ratios corresponding to this configuration (and thus to any configuration related to this one by a conformal transformation) are easily evaluated:
\be{crosrat}
u=\frac{x_{12}^2x_{34}^2}{x_{13}^2x_{24}^2}\Bigg\vert_{x_4\rightarrow\infty}=\vert z\vert^2 \ ,  \ \ \ \text{and analogously} \ \ \ v=\vert 1-z \vert^2 \ .
\ee

In what follows, important role will be played by the configuration having $z=1/2$. It can be mapped onto four points at the vertices of a square (both have $u=v=1/4$): 
\begin{center}
\includegraphics[height=3cm]{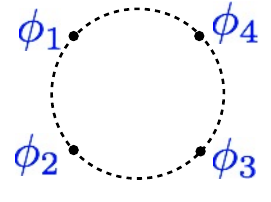}
\end{center}

The square configuration (and thus the $z=1/2$ configuration equivalent to it), and its neighborhood, will be most natural to study the conformal bootstrap equation. The reason is that this configuration treats symmetrically the (12)(34) with (14)(23) OPE channels, which are the channels compared in the bootstrap equation.

We will now introduce another configuration, which puts the points $x_1 \ \text{and} \ x_2$ (and $x_3$ and $x_4$) symmetrically with respect to the origin $x=0$, as in this figure:
\begin{center}
\includegraphics[height=5cm]{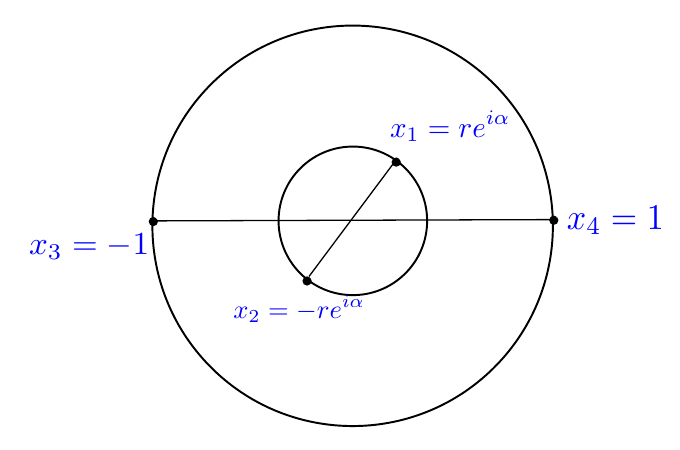}
\end{center}

This configuration is characterized by a complex parameter $\rho=r e^{i\alpha}$, $r<1$. It can of course be mapped by a conformal transformation onto the one characterized by $z$, for a certain value of $z$. To find the correspondence between $\rho$ and $z$ we just have to make sure that $u$ and $v$ are the same. An explicit computation shows that $u$ and $v$ will be the same if we fix
\be{mapzr}
z=\frac{4\rho}{(1+\rho)^2} \Leftrightarrow \rho=\frac{z}{(1+\sqrt{1-z})^2}.
\ee
E.g.~for $z=1/2$ we get $\rho=3-2\sqrt{2}\approx 0.17$.

\begin{figure}[h!]
\begin{center}
\includegraphics[scale=1]{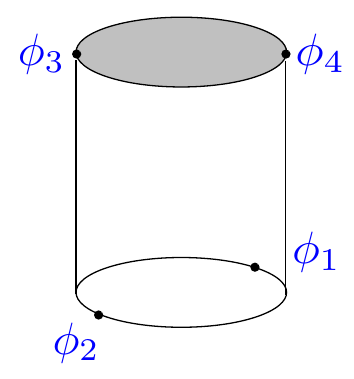}
\caption{The configuration on $S^{D-1}\times \bR$ obtained by a Weyl transformation. The pairs of points $\phi_{3,4}$ and $\phi_{1,2}$ are now in antipodal positions on the spheres at the cylinder time $0$ and $-\log r$. Their positions on these spheres are rotated with respect to each other by angle $\alpha$.}
\label{fig-cyl}
\end{center}
\end{figure}

As discussed in lecture \ref{lecture3}, we can apply a Weyl transformation to map the CFT dynamics from the flat space to the cylinder. The last configuration is then mapped to the one in Fig.~\ref{fig-cyl}. 

On the cylinder, conformal block of an exchanged operator $\calO_{\Delta,l}$ can be computed as:
\be{confblockcyl} 
\text{Conformal Block}=\sum \langle 0\vert\phi_1\phi_2\vert n\rangle e^{-E_n\tau}\langle n\vert \phi_3\phi_4\vert 0\rangle \ ,
\ee
where the sum is over all the descendants of $\vert\Delta,l\rangle$, $E_n=\Delta+n$, and $\tau=-\log r$ is the cylinder time interval along which we have to propagate the exchanged states. The product of the matrix elements depends only on $\alpha$, and so we conclude that the conformal block must have the form
\be{confblockcyl2}
\text{Conformal Block}=\sum_{n=0}^\infty A_n(\alpha)r^{\Delta+n} \ .
\ee
The coefficients $A_n$ are completely fixed by conformal symmetry. Their precise values can be found, but we will not pursue this problem here. The leading coefficient $A_0$ is easily determined on physical grounds. The states $\phi_1\phi_2\vert 0\rangle$ and $\phi_3\phi_4\vert 0\rangle$ differ by a rotation by angle $\alpha$. Therefore, $A_0(\alpha)$ measures how the matrix elements with a spin $l$ state $\vert\Delta,l\rangle$ change under the rotation by an angle $\alpha$. 

Let us parametrize the state on the cylinder by the unit vector $\vec n\in S^{D-1}$ pointing to the point where $\phi_1$ is inserted on the sphere. The state $\vert\Delta,l\rangle$ has internal indices $\vert\Delta,l\rangle_{\mu_1,\mu_2,\ldots}$ which form a symmetric traceless spin $l$ tensor. The individual matrix elements are:  
\be{exchrot}
\langle 0\vert\phi_1\phi_2\vert\Delta,l\rangle_{\mu_1,\mu_2,\ldots}=const.(\vec n_1^{\mu_1}\ldots\vec n_1^{\mu_l}-\text{traces}) \ ,
\ee
since there is only one traceless and symmetric spin $l$ tensor which can be constructed out of a single vector $\vec n_1$. 
Thus, up to normalization, the leading coefficient will be a contraction of two such matrix elements
\be{A01st}
A_0(\alpha)=(n_1^{\mu_1}\ldots\vec n_1^{\mu_l}-\text{traces})(n_2^{\mu_1}\ldots\vec n_2^{\mu_l}-\text{traces})=\mathcal P(\vec n_1.\vec n_2)=\mathcal P(\cos\alpha) \ ,
\ee
where $\mathcal P$ is a certain polynomial whose coefficients can only depend on the spin $l$ and on the number of spacetime dimensions $D$.

In $D=2$, symmetric traceless tensors have only two nonzero components 
$zz\ldots z \ \text{and} \ \bar z\bar z\ldots\bar z$, so
\be{2DA0}
A_0(\alpha)=(n_1^z n_2^{\bar z})^l + c.c.= \cos(l\alpha) \ .
\ee
For $D=3$, the answer is the Legendre polynomials $P_l(\cos\alpha)$
\be{3DA0}
A_0(\alpha)=P_l(\cos\alpha) \ .
\ee
 They are the same Legendre polynomials that appear in the wavefunction of the $m=0$ spin $l$ state. [{\bf Exercise:} Explain why this is not accidental.]  For $D=4$ we have
\be{4DA0}
A_0(\alpha)=\frac{1}{l+1}\frac{\sin[(l+1)\alpha]}{\sin\alpha} \ ,
\ee
the so-called Chebyshev polynomials. Generalization for any $D$ has the form:
\be{DDA0}
A_0(\alpha)=C_l^{(D/2-1)}(\cos\alpha) \ ,
\ee
where $C_l^{(D/2-1)}(\cos\alpha)$ are the Gegenbauer polynomials.

\begin{figure}[h!]
\begin{center}
\includegraphics[scale=0.5]{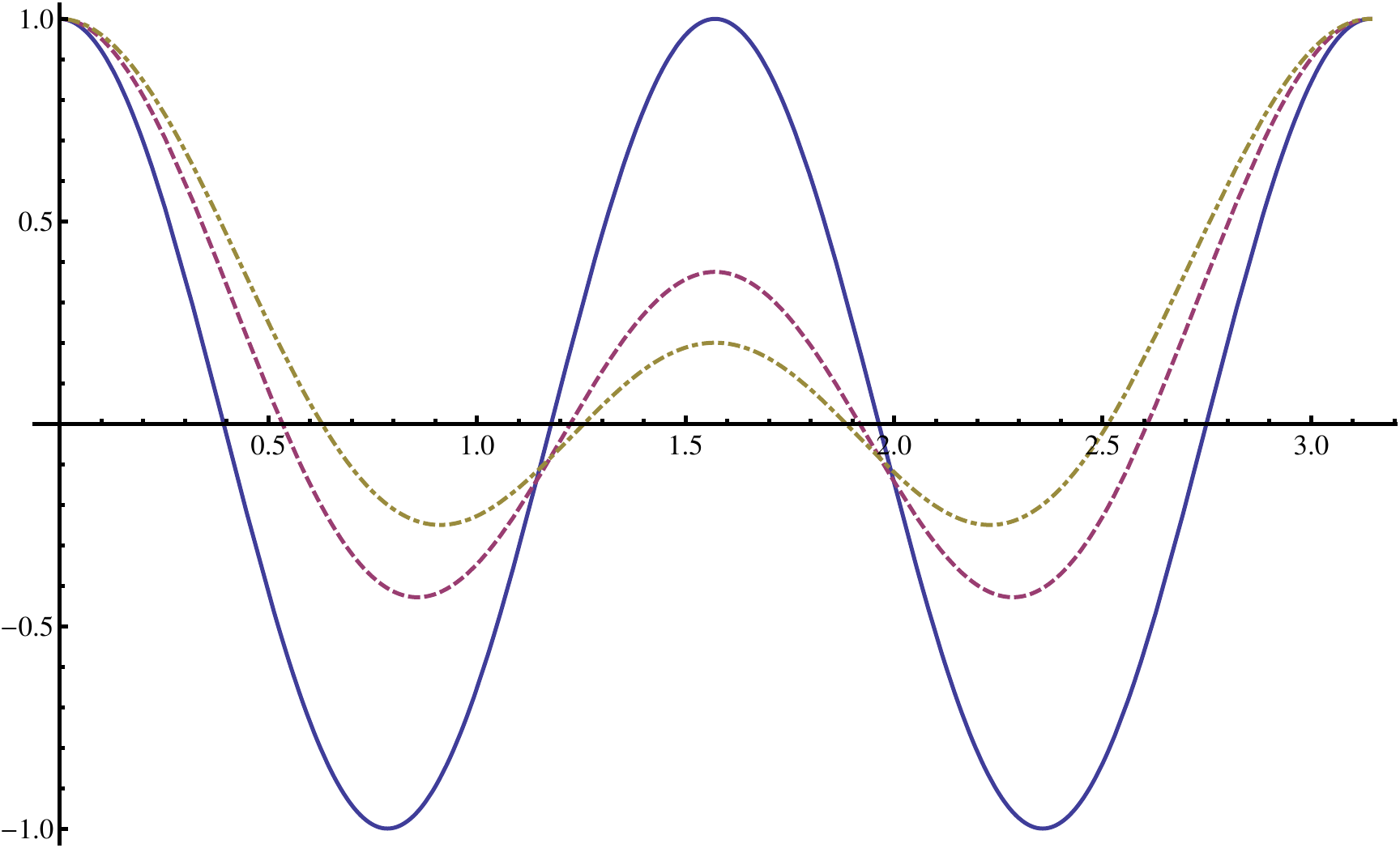}
\caption{$A_0(\alpha)$ for $l=4$ and $D=2,3,4$ shown by a solid, dashed, and dot-dashed curve.}
\label{poly}
\end{center}
\end{figure}

The functions $A_0(\alpha)$ for $l=4$ and different $D$ are plotted in Fig.~\ref{poly}. We normalize them so that $A_0(\alpha=0)=1$. After a bunch of oscillations, we get $A_0(\alpha=\pi)=(-1)^l$.

Note that given the small value of $r\approx 0.17$ for the configuration equivalent to $z=1/2$ and to the square configuration, the leading term in the expansion of the CB, 
\beq
A_0(\alpha)r^{\Delta},
\eeq 
constitutes a very good approximation near this point. The leading correction is $O(r^2)\sim$ few percent, since the term linear to the descendants will not appear if the operators are inserted symmetrically:
\be{aproxjust}
\phi(x)\phi(-x)\sim \mathcal O(0)+ c x^\mu x^\nu\partial_\mu\partial_\nu\mathcal O(0)+\ldots
\ee

\subsection{An application}
Now that we understood approximately the CB, let us look at the crossing symmetry equation. Consider the correlator of four identical scalar particles with dimension $\Delta_\phi=d$ (e.g.~they can be the spin fields of the Ising model). We have
\be{4pfising}
\langle \phi(x_1)\phi(x_2)\phi(x_3)\phi(x_4)\rangle=\frac{g(u,v)}{(x_{12})^{2d}(x_{34})^{2d}}\ ,
\ee 
with
\be{expguv}
g(u,v)=1+\sum_\mathcal O\lambda_\mathcal O^2G_\mathcal O(u,v) \ ,
\ee
where $1$ is the CB of the unit operator, and the structure of the $G_\mathcal O(u,v) $ is
\be{CBstructure}
G_\mathcal O(u,v)=C_l(\cos\alpha)r^\Delta[1+\mathcal O(r^2)] \ .
\ee
The relation between $u,v$ and $r,\alpha$ takes the form:
\be{prevuvar}
u=\vert z\vert^2\ , v=\vert 1-z\vert^2 \  \Rightarrow \rho=r e^{i\alpha}=\frac{z}{(1+\sqrt{1-z})^2}\ .
\ee

Moreover, in lecture \ref{lecture2}, Eq.~\reef{rat}, we saw that the crossing symmetry dictates
\be{dictcros}
v^dg(u,v)=u^dg(v,u) \ , \ \text{or} \ (v^d-u^d)+\sum_\mathcal O\lambda_\mathcal O^2[v^d G_\mathcal O(u,v)-u^dG_\mathcal O(v,u)]=0 \ .
\ee
This is what the bootstrap equation looks like in the case under consideration. It is simpler than in general, because the same operators occur in both OPE channels (12)(34) and (14)(23). So to compare the channels we can take the difference and collect the OPE coefficients, which gives \reef{dictcros}.

Still, this looks like a non-trivial equation since it is not satisfied term by term, but only in the sum---all operators work together. On the other hand, there is still a lot of freedom, as we can move the spectrum and adjust the $\lambda_\mathcal O$'s. 

A natural first question is as follows: What are the spectra for which we can find $\lambda_\mathcal O$'s such that the crossing is satisfied? Presumably, this will not happen for any spectrum. We will now show it rigorously,
assuming that the theory is unitary, so that all $\lambda_\calO^2>0$.

Eq.~\reef{dictcros} has to be satisfied for any $u$ and $v$, but for this simple demonstration let us look at the points having $0<z<1$ real, so that also $\rho$ is real ($\alpha=0$). Then we get
\be{simplcasa=0} 
[(1-z)^{2d}-z^{2d}]+\sum_\mathcal O\lambda_\mathcal O^2\{(1-z)^{2d}[\rho(z)]^\Delta-z^{2d}[\rho(1-z)]^\Delta\}=0 \ .
\ee
In this equation we replaced conformal blocks by their approximate expressions, which can be trusted near $z=1/2$ where both $\rho(z),\rho(1-z)\simeq 0.17$ and the omitted terms are suppressed by $0.17^2$. In what follows we will only work near $z=1/2$. 

Let us Taylor-expand near $z=1/2$. Only odd powers of $z-1/2$ will appear since the functions are odd around this point. From the first term we get
\be{expx}
[(1-z)^{2d}-z^{2d}]\simeq-C_d(x+\frac{4}{3}(d-1)(2d-1)x^3+\mathcal O(x^5)) \ ,
\ee
with $x=z-1/2$ and $C_d>0$ a positive constant. 

Now, suppose for the sake of the argument that all exchanged operators have $\Delta\ge \Delta_{min}$ with $\Delta_{min}\gg \max(d,1)$.\footnote{Note that in 2d the scalar unitarity bound is $d>0$.} We will now show that such an assumption is inconsistent. Expanding the conformal block terms in \reef{simplcasa=0}, we can neglect the variation of $z^{2d}$ and $(1-z)^{2d}$ factors (since $\Delta\gg d$). We get:
\be{Dbiggethand}
[\rho(z)]^\Delta-[\rho(1-z)]^\Delta\simeq B_\Delta \left(x+\frac{4}{3}\Delta^2x^3+\ldots\right) \ ,
\ee
where $B_\Delta>0$ is another positive constant. Let's change the conformal block normalization so that $B_\Delta=1$ (effectively incorporating this constant into $\lambda_\mathcal O^2$). Now let's require that \reef{simplcasa=0} be satisfied term by term in the Taylor expansion around $z=1/2$. We get:
\be{termbyterm}
\begin{aligned}
O(x):&-C_d+\sum_\mathcal O \lambda_\mathcal O^2=0\\
O(x^3):&-C_d \frac{4}{3}(d-1)(2d-1) +\frac{4}{3}\sum_\mathcal O\lambda_\mathcal O^2\Delta^2 =0 \ .
\end{aligned}
\ee
The $O(x^5)$ terms etc would give more equations but we won't need them for now.

We have the following chain of equations:
\be{consfail}
\Delta_{min}^2\underbrace{\sum_\mathcal O\lambda_\mathcal O^2}_{=C_d}\le\sum_\mathcal O\lambda_\mathcal O^2\Delta^2=(d-1)(2d-1)C_d \ ,
\ee
where in the first term we used the $O(x)$ expansion equation, the inequality is true since all $\lambda_\calO^2>0$, and the last equality is a consequence of the $O(x^3)$ expansion relation. 
So we conclude:
\be{contradic}
\Delta_{min}\le \sqrt{(d-1)(2d-1)} \ ,
\ee
which is a contradiction, since we started by assuming $\Delta_{min}\gg \max(d,1)$.

We arrive at the following conclusion. The crossing symmetry, together with unitarity, requires that the OPE contain operators of low dimension 
\be{crOPEreq}
\Delta_{min}\le f(d) \ .
\ee
The above method gives a very rough estimate for $f(d)$. But clearly, the analysis can be improved:

1. Relax the condition $z=\bar z$, in order to be able to distinguish between operators of different spins. 

2. Use exact expression for the CBs.

3. Expand to higher order in $x=z-1/2$, $\bar x=\bar z-1/2$.

\begin{figure}[htbp]
\begin{center}
\includegraphics[scale=0.6]{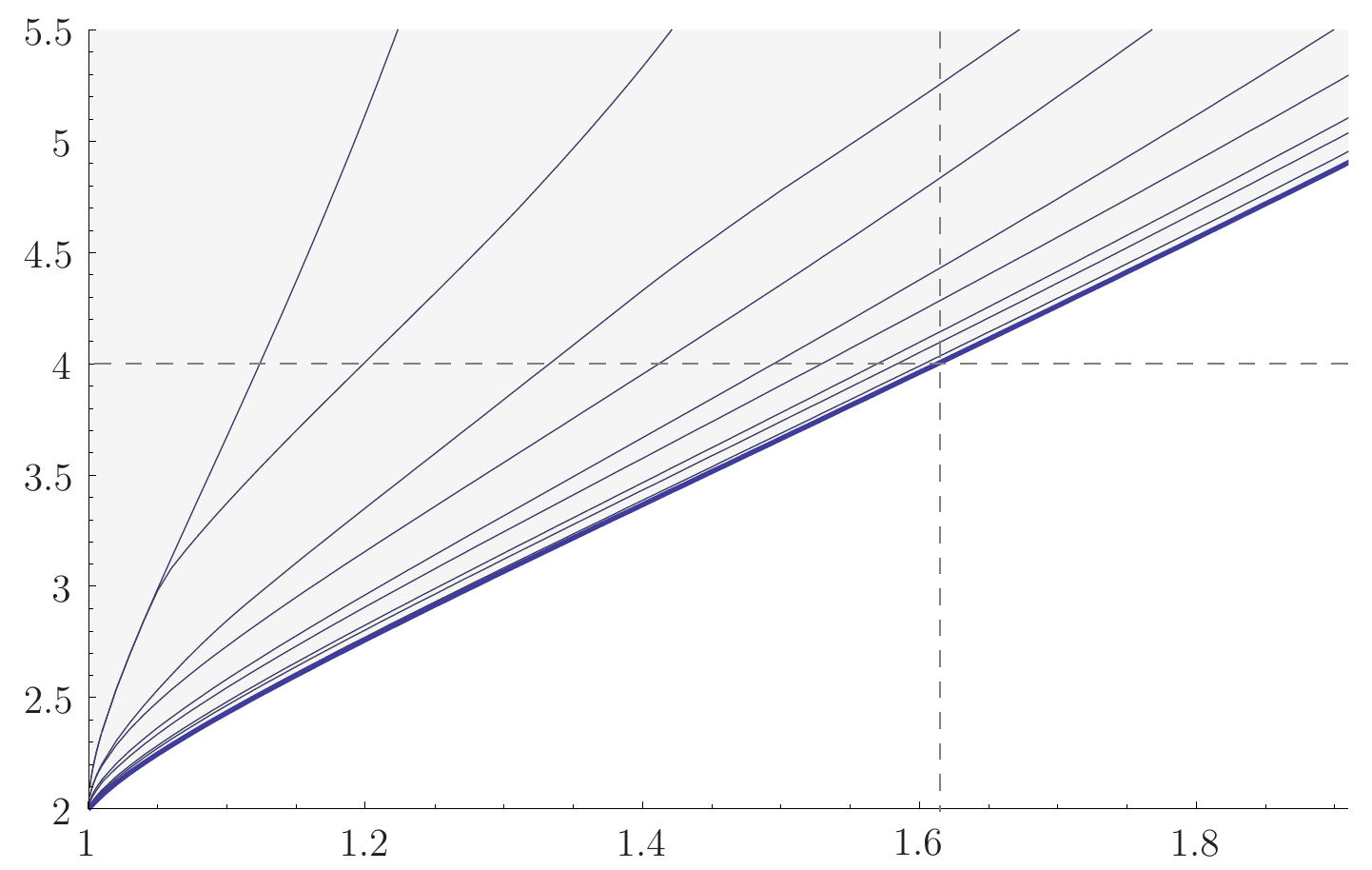}
\caption{Horizontal axis: dimension of a scalar primary $\phi$ in a $D=4$ CFT. Vertical axis: The upper bound on the dimension of the first scalar appearing in the OPE $\phi\times\phi$. The gray area is thus excluded. The various curves correspond to expanding to higher and higher order; the bound eventually converges.}
\label{carving}
\end{center}
\end{figure}

By doing that, one can find the bound in $D=4$ dimensions shown in Fig. \ref{carving} (the original bound is due to Rattazzi, Rychkov, Tonni, Vichi \cite{Rattazzi:2008pe}; the shown plot is from a paper by Poland, Simmons-Duffin and Vichi \cite{Poland:2011ey})

This is where these notes end. For further information about the bootstrap in $D\ge 3$ you will have to turn to the lectures mentioned in the introduction, and to the original papers mentioned below.

\section*{Literature}

\emph{Early history of conformal bootstrap} 

In the early 1970s, there were three groups of people working on CFTs.
In Italy, there was a group of Raoul Gatto who collaborated with Sergio Ferrara, Aurelio Grillo and Giorgio Parisi.
In Germany, there were Gerhard Mack and Martin L\"uscher. In Russia, Alexander Polyakov and Alexander Migdal. The three original papers on conformal bootstrap came from these groups: Ferrara, Grillo, and Gatto \cite{Ferrara:1973yt}, Polyakov \cite{Polyakov:1974gs}, and Mack \cite{Mack:1975jr}. 

It's interesting to compare these early works. Gatto's group wrote a long series of CFT papers in 1971-75,
with many important results: conformally invariant OPE, conformal blocks, the so-called ``shadow formalism", etc. Their paper \cite{Ferrara:1973yt} contains the crossing symmetry constraint (their Eq.~(6.33)), although they stopped short of claiming that it should be enough to fix the theory.

Polyakov's paper \cite{Polyakov:1974gs} starts with a programmatic tour-de-force promoting the crossing symmetry constraints to the dynamical principle which should be used to solve the theory. The precise implementation of these constraints developed in later sections of the paper is different from the one used nowadays: instead of conformal blocks, Polyakov advocates the use of ``unitary blocks"---objects that he defines and which violate the OPE by logarithmic terms. His constraint on the spectrum and the OPE coefficients is that these violating terms must cancel when summed over all fields appearing in the OPE. He works out an example in $4-\eps$ dimensions reproducing the known lowest-order anomalous dimensions. 

Although Polyakov's philosophy forms the basis of today's bootstrap, his alternative bootstrap equation remains virtually unknown and unexplored. Recently, Sen and Sinha \cite{Sen:2015doa} reproduced Polyakov's computation and extended it one order higher. Hopefully, it will lead to rapid further progress in this direction.

Mack's paper \cite{Mack:1975jr} appeared three years later than \cite{Ferrara:1973yt,Polyakov:1974gs}; it cites both other papers. It stresses that OPE in CFTs gives a convergent expansion at finite point separation, as rigorously proved by Mack himself \cite{Mack:1976pa} around the same time.

The early interest in CFT was motivated by the search for a theory of strong interactions at high energies and, especially for the Russians, by connections to the theory of critical phenomena. 
In mid-70's competing theories emerged which allowed rapid progress on both of these problems: QCD for the strong interactions, and Wilson's RG for the critical indices. CFT no longer seemed so urgent or relevant, and it entered a hibernation period\ldots 

\ldots from which it emerged triumphantly in 1984 with the famous paper by Belavin, Polyakov and Zamolodchikov \cite{Belavin:1984vu} developing CFT in $d=2$. Actually, it's this paper which proposed the term ``conformal bootstrap" to denote the study of crossing symmetry constraints expressed in terms of conformal blocks. Before that, Migdal called ``conformal bootstrap" a hybrid technique in which Feynman-diagrammatic expansion in the IR region was computed by means of resummed vertices and propagators respecting conformal invariance, and a Schwinger-Dyson equation was imposed to fix the expansion parameter. The second usage of the term still occurs occasionally, creating a minor confusion.

\emph{Other references}
  
For the unitarity bounds in 2d, see the yellow book \cite{yellow}.

For the conformal blocks in $D\ge 3$, start with the very influential papers of Dolan and Osborn \cite{DO1,DO2}.
See our work \cite{Costa:2011dw} for an introduction to the issues arising for the external fields with spin, and \cite{Hogervorst:2013sma} for the expansion in Gegenbauer polynomials. David Simmons-Duffin work \cite{SimmonsDuffin:2012uy} provides a modern introduction to the ``shadow formalism". The recursion relation by Kos, Poland and Simmons-Duffin \cite{Kos:2013tga}, section 3, is the state-of-the-art tool for computing conformal blocks for the external scalar operators in an arbitrary $D$. The detailed knowledge of conformal blocks is a prerequisite for any bootstrap analysis, and their study is a rapidly developing subject, so the above is only a very partial list of references.
  
Conformal bootstrap bounds were first discovered, in $D=4$, in our paper \cite{Rattazzi:2008pe}. Interestingly, at the time we were interested in a sharply formulated question of electroweak phenomenology beyond the Standard Model, and conformal bootstrap was just a tool to answer that question. Since then all sorts of bounds were derived, across theories in different $D$ and with different (super)symmetry assumptions.

Remarkably the bounds often exhibit ``kinks", and various famous theories were conjectured to live at those kinks. E.g.~this seems to be true for the critical point of the 3d Ising model \cite{ElShowk:2012ht,El-Showk:2014dwa}, giving a very precise determination of its low-lying spectrum (subject to the kink hypothesis).

An exciting recent development is that bounds sometimes reduce to tiny islands when several 4pt functions are studied together. This happens for the 3d Ising model \cite{Kos:2014bka,Simmons-Duffin:2015qma} and for the $O(N)$ models \cite{Kos:2015mba}. This method gives the world's most precise, and rigorous, determination of the 3d Ising model critical exponents.

To finish this micro-review, we would like to mention two lines of research which are complementary to deriving rigorous bounds. 

1. Gliozzi and collaborators \cite{Gliozzi:2013ysa,Gliozzi:2014jsa,Gliozzi:2015qsa} developed a ``severe truncation" method in which only a handful of low-lying operators are kept in the conformal block expansion. Imposing a judiciously chosen subset of the infinitely many bootstrap equations (in the usual derivative expansion around $z=\bar z=1/2$), one sometimes manages to fix the scaling dimensions with remarkable precision. This method is not yet as systematic as the bounds method. It's worth developing it further, also because it applies to both unitary and non-unitary theories.

2. Expansion around $z=\bar z=1/2$ is adequate if one wants to constraint operators of low dimensions. Another privileged class of operators are those of low \emph{twist} $\tau=\Delta-l$. These operators dominate the OPE when one approaches the Minkowski light cone limit $z\to0$, $\bar z$ fixed (we have not explained this but in Minkowski space $z,\bar z$ become independent real parameters). Interestingly, applying the bootstrap equations near the light cone, one can get rigorous asymptotic results about the spectrum of operators with $\tau=O(1)$ and $l\to\infty$ \cite{Fitzpatrick:2012yx,Komargodski:2012ek}. It would be nice to find a way to combine these analytic results with the numerical results about the low-dimension spectrum, to get a picture of the Minkowski 4pt function in the full region $0<z,\bar z<1$ where it is regular. For recent work in this direction, see \cite{Alday:2015ota,Alday:2015ewa}.

\bibliography{EPFL-Biblio}
\bibliographystyle{utphys.bst}
  
\end{document}